\documentclass[runningheads]{llncs}
\usepackage[T1]{fontenc}
\usepackage{geometry}
\usepackage{amsmath}
\usepackage{graphicx}
\usepackage{enumerate}
\usepackage{booktabs} 
\usepackage{url} 
\usepackage{subcaption}
\usepackage{lipsum}
\usepackage{multirow}
\usepackage{xcolor}
\usepackage{lineno}
\usepackage{makecell}
\usepackage{algorithm}
\usepackage{algorithmic}
\usepackage{booktabs}
\usepackage{wasysym}
\usepackage{amssymb}
\usepackage{pifont}
\usepackage{caption}
\usepackage[export]{adjustbox}
\usepackage{hyperref} 
\usepackage{bm}
\usepackage{array}
\usepackage[figuresright]{rotating}
\usepackage[misc]{ifsym}
\usepackage{cite}
\usepackage{CJKutf8}
\usepackage{setspace}
\usepackage{capt-of}
\usepackage{placeins} 
\usepackage{longtable}
\usepackage{tabularx}
\usepackage{ragged2e}

\makeatletter
\def\UrlAlphabet{%
      \do\a\do\b\do\c\do\d\do\e\do\f\do\g\do\h\do\i\do\j%
      \do\k\do\l\do\m\do\n\do\o\do\p\do\q\do\r\do\s\do\t%
      \do\u\do\v\do\w\do\x\do\y\do\z\do\A\do\B\do\C\do\D%
      \do\E\do\F\do\G\do\H\do\I\do\J\do\K\do\L\do\M\do\N%
      \do\O\do\P\do\Q\do\R\do\S\do\T\do\U\do\V\do\W\do\X%
      \do\Y\do\Z}
\def\UrlDigits{\do\1\do\2\do\3\do\4\do\5\do\6\do\7\do\8\do\9\do\0}
\g@addto@macro{\UrlBreaks}{\UrlOrds}
\g@addto@macro{\UrlBreaks}{\UrlAlphabet}
\g@addto@macro{\UrlBreaks}{\UrlDigits}
\makeatother

\usepackage{tikz}
\newcommand*{\circled}[1]{\lower.7ex\hbox{\tikz\draw (0pt, 0pt)%
    circle (.5em) node {\makebox[1em][c]{\small #1}};}}

\begin{document}
\title{Enhancing Diagnostic Accuracy in Rare and Common Fundus Diseases with a Knowledge-Rich Vision-Language Model}
\author{Meng Wang \inst{1,2\#}\and
Tian Lin\inst{3 \#}\and
Aidi Lin\inst{3}\and
Kai Yu\inst{4}\and
Yuanyuan Peng\inst{5}\and
Lianyu Wang\inst{6}\and
Cheng Chen\inst{7}\and
Ke Zou\inst{8}\and \\
Huiyu Liang\inst{3}\and
Man Chen \inst{3}\and
Xue Yao\inst{3}\and
Meiqin Zhang\inst{3}\and
Binwei Huang\inst{3}\and
Chaoxin Zheng\inst{3}\and
Peixin Zhang \inst{3}\and
Wei Chen\inst{3}\and \\
Yilong Luo\inst{3}\and 
Yifan Chen\inst{3}\and
Honghe Xia \inst{3}\and
Tingkun Shi \inst{3}\and
Qi Zhang \inst{3}\and
Jinming Guo \inst{3}\and
Xiaolin Chen \inst{3}\and 
Jingcheng Wang\inst{9}\and\\ 
Yih Chung Tham\inst{1,2}\and
Dianbo Liu \inst{1,2}\and
Wendy Wong \inst{1,2}\and
Sahil Thakur \inst{10}\and
Beau Fenner \inst{10,11}\and
Danqi Fang \inst{12}\and 
Siying Liu \inst{13}\and \\
Qingyun Liu \inst{13}\and 
Yuqiang Huang \inst{3}\and 
Hongqiang Zeng \inst{14}\and 
Yanda Meng \inst{15}\and
Yukun Zhou \inst{16,17,18}\and
Zehua Jiang \inst{19,20}\and \\
Minghui Qiu \inst{21}\and
Changqing Zhang \inst{22}\and
Xinjian Chen \inst{23}\and
Sophia Y. Wang \inst{24}\and
Cecilia S. Lee \inst{25,26}\and \\
Lucia Sobrin \inst{27}\and
Carol Y Cheung \inst{12}\and
Chi Pui Pang \inst{2, 12}\and
Pearse A. Keane\inst{17,18}\and \\
Ching-Yu Cheng\inst{1,2,10,11 (\textrm{\Letter})}\and
Haoyu Chen\inst{3 (\textrm{\Letter})}\and
Huazhu Fu\inst{28 (\textrm{\Letter})}
}
\titlerunning{RetiZero}
\authorrunning{M. Wang et al.}
%
\institute{
Centre for Innovation \& Precision Eye Health, Yong Loo Lin School of Medicine, National University of Singapore, Singapore 117549, Singapore. \and
Department of Ophthalmology, Yong Loo Lin School of Medicine, National University of Singapore, Singapore 117549, Singapore. \and
Joint Shantou International Eye Center, Shantou University and the Chinese University of Hong Kong, 515041 Shantou, Guangdong, China. \and
Department of Radiology, University of Pennsylvania, Philadelphia, PA 19104, USA. \and
School of Biomedical Engineering, Anhui Medical University, 230032 Hefei, Anhui, China.\and
College of Computer Science and Technology, Nanjing University of Aeronautics and Astronautics, 211100 Nanjing, Jiangsu, China.\and
Center of Advanced Medical Computing and Analysis, Massachusetts General Hospital and Harvard Medical School, Boston, MA 02114, USA.\and
National Key Laboratory of Fundamental Science on Synthetic Vision and the College of Computer Science, Sichuan University, 610065 Chengdu, Sichuan, China.\and
Big Vision Medical Technology Ltd., Suzhou, China.\and
Singapore Eye Research Institute, Singapore National Eye Centre, Republic of Singapore.\and
Ophthalmology \& Visual Sciences Academic Clinical Program (EYE ACP), Duke-NUS Medical School, Singapore.\and
Department of Ophthalmology and Visual Sciences, The Chinese University of Hong Kong, 999077 Hong Kong, China.\and
Shenzhen Longgang E.N.T Hospital, 518172, Shenzhen, Guangdong, China. \and
Dongguan Songshan Lake Central Hospital, 523326, Dongguan, Guangdong, China. \and
Department of Computer Science, University of Exeter, Exeter, EX4 4RN, UK.\and
Centre for Medical Image Computing, University College London, London, UK. \and 
NIHR Biomedical Research Centre at Moorfields Eye Hospital NHS Foundation Trust, London, UK. \and 
Institute of Ophthalmology, University College London, London, UK. \and
Tsinghua Medicine of Tsinghua University, 100084, Beijing, China.\and
School of Clinical Medicine, Beijing Tsinghua Changgung Hospital, 102218, Beijing, China.\and
Foshan Aier Eye Hospital, 528000, Foshan, Guangdong, China.\and
College of Intelligence and Computing, Tianjin University, 300350 Tianjin, China.\and
School of Electronics and Information Engineering, Soochow University, Jiangsu 215006, China.\and
Byers Eye Institute, Department of Ophthalmology, Stanford University School of Medicine, Palo Alto, California, USA.\and
Department of Ophthalmology, University of Washington, Seattle, WA, USA.\and
Roger H. and Angie Karalis Johnson Retina Center, Seattle, WA, USA.\and
Department of Ophthalmology, Massachusetts Eye and Ear, Harvard Medical School, Boston, MA, USA.\and
Institute of High Performance Computing (IHPC), Agency for Science, Technology and Research (A*STAR), 1 Fusionopolis Way, \#16-16 Connexis, Singapore 138632, Republic of Singapore.\\
\# M. Wang, and T. Lin are the co-first authors.\\
\textrm{\Letter} C.Y. Cheng, H. Chen, and H. Fu are the co-corresponding authors and contributed equally.
}
\captionsetup[figure]{labelfont={bf},name={Figure: },labelsep=period}
\captionsetup[table]{labelfont={bf},name={Table: },labelsep=period}
\maketitle              
\begin{abstract}
Previous foundation models for fundus images were pre-trained with limited disease categories and knowledge base. Here we introduce a knowledge-rich vision-language model (RetiZero) that leverages knowledge from more than 400 fundus diseases. For RetiZero’s pre-training, we compiled 341,896 fundus images paired with texts, sourced from public datasets, ophthalmic literature, and online resources, encompassing a diverse range of diseases across multiple ethnicities and countries. RetiZero exhibits remarkable performance in several downstream tasks, including zero-shot disease recognition, image-to-image retrieval, AI-assisted clinical diagnosis, few-shot fine-tuning, and internal- and cross-domain disease identification. In zero-shot scenarios, RetiZero achieves Top-5 accuracies of 0.843 for 15 diseases and 0.756 for 52 diseases. For image retrieval, it achieves Top-5 scores of 0.950 and 0.886 for the same sets, respectively. AI-assisted clinical diagnosis results show that RetiZero’s Top-3 zero-shot performance surpasses the average of 19 ophthalmologists from Singapore, China, and the United States. RetiZero substantially enhances clinicians’ accuracy in diagnosing fundus diseases, in particularly rare ones. These findings underscore the value of integrating the RetiZero into clinical settings, where various fundus diseases are encountered. 

\end{abstract}
\newpage
\section{Introduction}
\noindent 
Blindness and visual impairment represent a substantial disease burden globally, impacting millions of individuals across all populations. Detection and timely treatment of ocular conditions, such as retinal and optic nerve diseases, are crucial for reducing severe and permanent damage. However, the insufficient availability of ophthalmic medical resources severely limits the prompt screening and management of fundus diseases with vast regional differences in many parts of the world.

In recent years, artificial intelligence (AI)-based fundus disease screening systems have been proposed and achieved promising performance on fundus disease detection and patients’ referral. Nevertheless, most previous AI-based methods were customized for and limited to specific diseases, such as diabetic retinopathy (DR)~\cite{bellemo2019artificial,xie2020artificial}, glaucoma~\cite{liu2019development,wang2020characterization}, and retinopathy of prematurity~\cite{peng2021automatic,taylor2019monitoring}. Although several methods were proposed for simultaneously screening multiple fundus diseases with promising performance~\cite{de2018clinically,wang2023uncertainty,cen2021automatic}, most current AI models for ocular disease screening were trained on task-specific datasets, leading to inevitable errors in detection when there were new data (e.g., images acquired by different camera) or changes in tasks (e.g., introducing new or rare categories). Furthermore, due to limited healthcare resources and the varying prevalence of fundus disease, collecting comprehensive datasets covering all kinds of fundus abnormalities is time-consuming and challenging. Consequently, most AI models were trained on limited data and disease categories, restricting their feature representation. Applying these models to different real-world settings or tasks requires extensive retraining with large datasets. Moreover, data quality and labeling issues further limit the widespread adoption of AI models in ophthalmic clinical settings, especially from a global perspective.

Driven by the abundance of big data and robust computing hardware, large foundation models (LFMs) have excelled in computer vision tasks~\cite{huang2023visual,radford2021learning}. Pre-trained on massive datasets, LFMs provide rich feature support for downstream tasks, such as object detection~\cite{zhang2020asymmetric}, few-shot recognition~\cite{zhang2022few}, and zero-shot~\cite{lai2024carzero}, etc. The first ophthalmic LFM, RETFound~\cite{RETFound}, introduced in 2023, was trained on large, unannotated retinal images using a masked autoencoder (MAE) framework~\cite{he2022masked}. It provides rich feature support and improves the performance of downstream tasks. However, such an approach can hinder the model's capacity to align feature information with labels in downstream tasks. In contrast, the Foundation LAnguage-Image model of the Retina (FLAIR)~\cite{FLAIR}, a Contrastive Language-Image Pre-training (CLIP)-based LFMs enhance feature representation by aligning text descriptions with image features, improving feature-label alignment but having difficulties with complex semantic features in medical imaging~\cite{huang2023visual}. MAE-based pretraining approaches excel in capturing complex semantic features in medical imaging by leveraging masked autoencoding techniques that focus on reconstructing obscured regions of an image, thereby fostering a deep understanding of local structures and subtle pathological details essential for accurate diagnostics. This approach encourages the model to learn rich, fine-grained representations by emphasizing contextual and structural information within the image. In contrast, CLIP-based pretraining primarily optimizes for global image-text alignment, aligning entire images with their corresponding textual descriptions without delving into the intricate internal features. While CLIP models are effective for tasks requiring broad semantic understanding and cross-modal associations, their emphasis on image-level alignment limits their ability to discern and interpret the nuanced and complex semantic patterns crucial in medical contexts. Consequently, CLIP-based methods struggle to effectively handle the detailed and sophisticated features necessary for precise medical image analysis, suggesting a significant limitation in their use for healthcare diagnostics. Furthermore, current LFMs for ophthalmic imaging are pre-trained on extensive yet categorically limited datasets. Therefore, developing LFMs with comprehensive ophthalmic disease knowledge would be crucial for representing complex retinal features to enhance downstream task performance. Nevertheless, collecting massive and diverse ophthalmic data that covers a wide range of fundus diseases for pretraining remains a significant challenge.

To address these problems and challenges, we collected 341,896 fundus images-text pairs from 29 publicly available datasets (containing 303,124 fundus images with labels), 180 ophthalmic literature (23,328 fundus images with diseases-related keywords), and online resources (15,544 fundus image-text pairs), encompassing over 400 retinal and optic nerve diseases across multiple countries, regions and ethnicities (Fig.1a). As shown in Fig. 1b, our LFM, RetiZero, is based on a contrastive vision-language pretraining framework that integrates MAE-based pretraining knowledge and low-rank training methods. Moreover, we introduced an uncertainty vision-language feature calibration method using Dirichlet reparameterization within the contrastive vision-language pretraining framework, to further align vision and language features in the high-dimensional embedding space. Consequently, RetiZero achieved superior performance in various downstream tasks, including zero-shot fundus disease recognition, image-to-image fundus disease retrieval, AI-assisted clinical diagnosis, internal domain fundus disease identification, few-shot fine-tuning, and cross-domain fundus disease identification.

\begin{figure}[t]
 \begin{center}
  \includegraphics[width=1\textwidth, height=1\textheight, keepaspectratio]{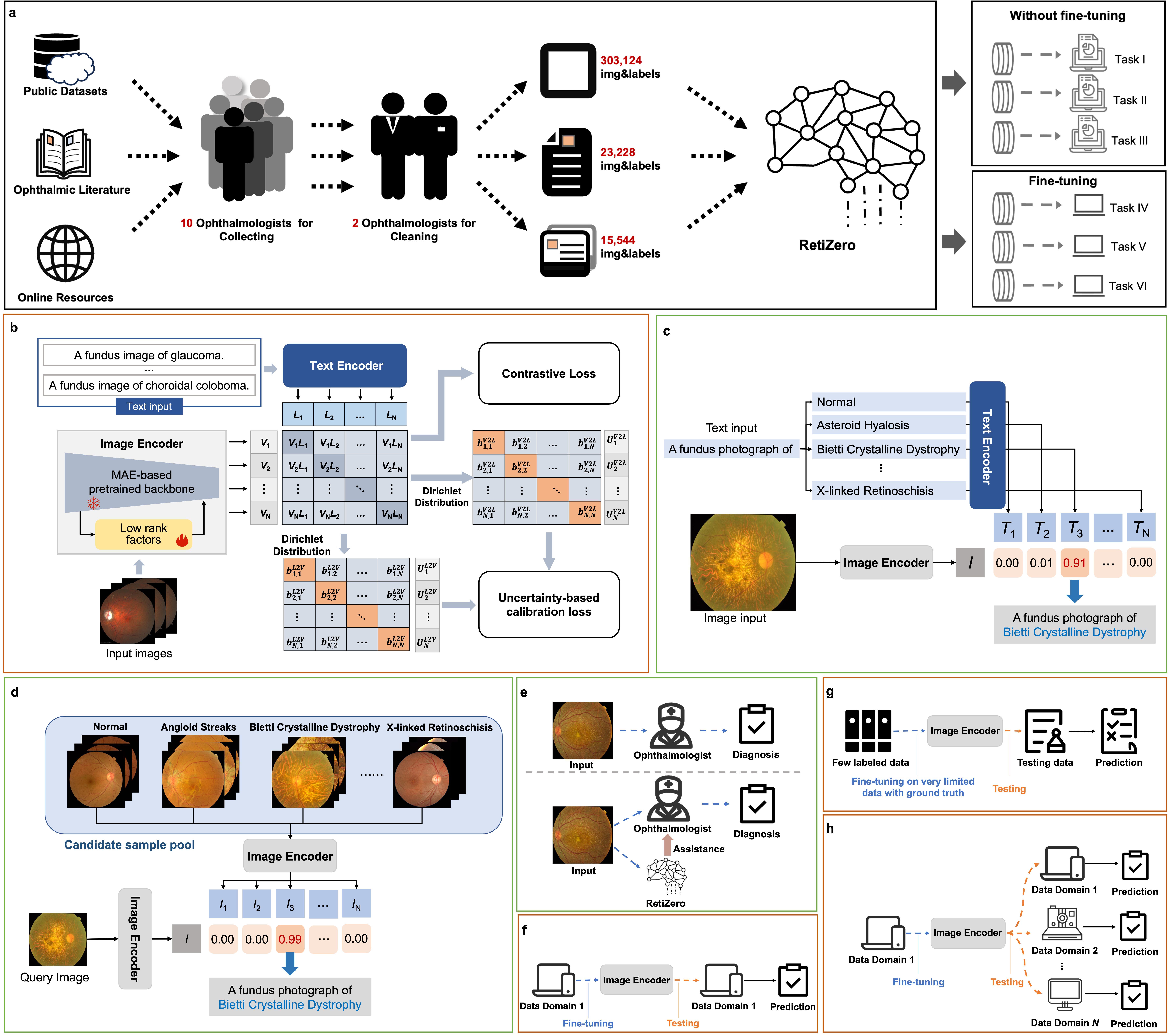}
 \end{center}
 \caption{\textbf{Overview of the framework.} a, Datasets for RetiZero pretraining: The RetiZero model was pre-trained using data from three primary sources: public datasets, ophthalmic literature, and online resources. We assembled a team of 12 ophthalmologists for manual data collection and cleaning. This involved downloading images and corresponding labels from public datasets, extracting images and corresponding disease-related keywords from ophthalmic literature, and downloading retinal diseases-relevant image-text pairs from online resources. b, RetiZero, which combines the strengths of self-supervised learning based on the MAE architecture and contrastive learning from the CLIP architecture. Moreover, we introduce an uncertainty vision-language feature calibration method into the contrastive vision-language pretraining framework, to further calibrate visual-language features in the high-dimensional embedding space. c, Task I: Zero-shot fundus disease recognition. d, Task II: Fundus disease identification by image-to-image retrieval. e, Task III: AI-assisted clinical diagnosis. f, Task IV: Internal domain retinal disease identification. "Internal domain" means that we fine-tuned and tested the model using the data with similar feature distribution. g, Task V: Few-shot fine-tuning. We evaluate RetiZero's performance in identifying fundus diseases with very limited training data. h, Task VI: Cross-domain fundus disease identification. "Cross-domain" means that we fine-tuned and tested the model using the data with different feature distributions.}
 \label{Overview}
\end{figure}
\FloatBarrier

\section{Results}
\subsection{Zero-shot fundus disease recognition}
The biggest advantage of RetiZero is the capability of zero-shot learning, which enables RetiZero to recognize fundus diseases using only textual prompts, without needing to retrain or fine-tune the model with labelled fundus images (Task I in Fig. 1c). As shown in Fig. 2a, RetiZero achieved overall Top-1, Top-3, and Top-5 scores of 0.442, 0.702, 0.840, respectively, for recognizing 15 common fundus diseases and normal condition of 30,089 fundus images (Eye-15 dataset, see in Supplementary Table 1). These scores of RetiZero improved by 25.5\% for Top-1 (P = 0.01), 15.7\% for Top-3 (P = 0.03), and 15.6\% for Top-5 (P = 0.04) over FLAIR. Furthermore, in the analysis of individual diseases, RetiZero showed remarkable zero-shot capability in identifying most categories, especially for retinitis pigmentosa (Top-1: 0.819, Top-3: 0.953, and Top-5: 0.974), retinal detachment (Top-1: 0.762, Top-3: 0.906, and Top-5: 0.963), and glaucoma (Top-1: 0.748, Top-3: 0.929, and Top-5: 0.972) (Extended Data Fig. 1).  To further validate RetiZero's zero-shot capability in more challenging clinical scenarios, we assembled a more demanding dataset named EYE-52 (see in Supplementary Table 2). This dataset included 7,007 fundus images from various ophthalmology clinics, covering 52 fundus diseases. Many of these diseases are rare in eye clinics but can lead to severe visual impairment if left undiagnosed. The incidence and prevalence of each category in the EYE-52 dataset were shown in Supplementary Table 3. As depicted in Fig. 2b, RetiZero achieved overall Top-1, Top-3, and Top-5 scores of 0.360, 0.626, and 0.756, respectively, for recognizing these 52 types of fundus diseases in a zero-shot manner, providing superior performance compared FLAIR (0.092, 0.263, and 0.340, respectively) and Random recognizing (0.029, 0.088, and 0.147, respectively). Furthermore, RetiZero demonstrated superior zero-shot performance, especially for recognizing several rare fundus diseases. For instance, RetiZero achieved Top-1, Top-3, and Top-5 scores of 0.616, 0.791, and 0.861, respectively, for identifying Bietti Crystalline dystrophy; and 0.509, 0.808, and 0.915, respectively for recognizing chorioretinal coloboma, the Top-1, Top-3, and Top-5 scores were (Extended Data Fig. 2). Fig. 2c shows the Top-5 prediction results provided by RetiZero and FLAIR for three rare fundus disease samples. More details on the rest of the 52 disease categories can be found in Extended Data Fig. 2. 
\subsection{Fundus disease identification by image-to-image retrieval}
For image-to-image retrieval (Fig.1d, TaskII), we treated each fundus image in turn as the query, removing it from the main set to form the candidate pool of all remaining images. We then passed the query through RetiZero’s image encoder, generating a feature embedding. Using the same encoder, we extracted embeddings for every image in the candidate pool and computed similarity scores between the query’s embedding and each candidate’s embedding. These similarity scores ranked the candidate images by how closely they matched the query, allowing us to retrieve the Top-K matches (e.g., Top-5). Fig. 2d and Extended Data Fig. 3 illustrate the excellent performance of RetiZero in identifying 15 fundus diseases through image-to-image retrieval. The overall scores for Top-1, Top-3, and Top-5 are 0.854, 0.928, and 0.950, respectively, representing an improvement of 9.4\%, 4.8\%, and 3.2\% over RETFound (all P < 0.001), and 300.2\%, 121.1\%, 74.0\% over FLAIR (all P < 0.001). In addition, RetiZero demonstrated the best performance across all categories compared to RETFound and FLAIR (Extended Data Fig. 3). In the more challenging Eye-52 dataset, RetiZero achieved overall Top-1, Top-3, and Top-5 scores of 0.726, 0.843, and 0.886, respectively (Fig. 2e). These scores represented improvements of 12.4\%, 7.9\%, and 6.3\% over RETFound (all P < 0.001), and 767.8\%, 389.9\%, 271.5\% over FLAIR, respectively (all P < 0.001). Fig. 2f shows an example of the Top-5 prediction results from RetiZero, RETFound, and FLAIR. Furthermore, in the analysis of individual diseases, RetiZero demonstrated remarkable potential, particularly in identifying several rare fundus diseases, such as punctate inner choroidopathy multifocal choroiditis (Top-1: 0.902, Top-3: 0.946, and Top-5: 0.962), chorioretinal coloboma (Top-1: 0.819, Top-3: 0.893, and Top-5: 0.910), and Bietti crystalline dystrophy (Top-1: 0.861, Top-3: 0.936, and Top-5: 0.942). More details on the 52 disease categories can be found in Extended Data Fig. 4. In addition, we calculated Precision@1, Precision@3, and Precision@5 to comprehensively evaluate RetiZero's performance in the task of fundus disease identification through image-to-image retrieval. Fig. 2d and Fig. 2e demonstrate that RetiZero achieved the highest Precision@1, Precision@3, and Precision@5 on both testing datasets, EYE-15 and EYE-52. Meanwhile, RetiZero demonstrated the best performance across most categories compared to RETFound and FLAIR (Extended Data Fig. 5 and Extended Data Fig. 6). Furthermore, we further visualized the heatmaps of different models' weights for various fundus diseases using GradCAM~\cite{selvaraju2017grad}. Extended Data Fig. 7 presents heatmaps illustrating the weights of different foundational models for various fundus diseases. RetiZero's weights were more precisely concentrated on the regions affected by different fundus diseases.

\begin{figure}[h]
 \begin{center}
  \includegraphics[width=1\textwidth, height=1\textheight, keepaspectratio]{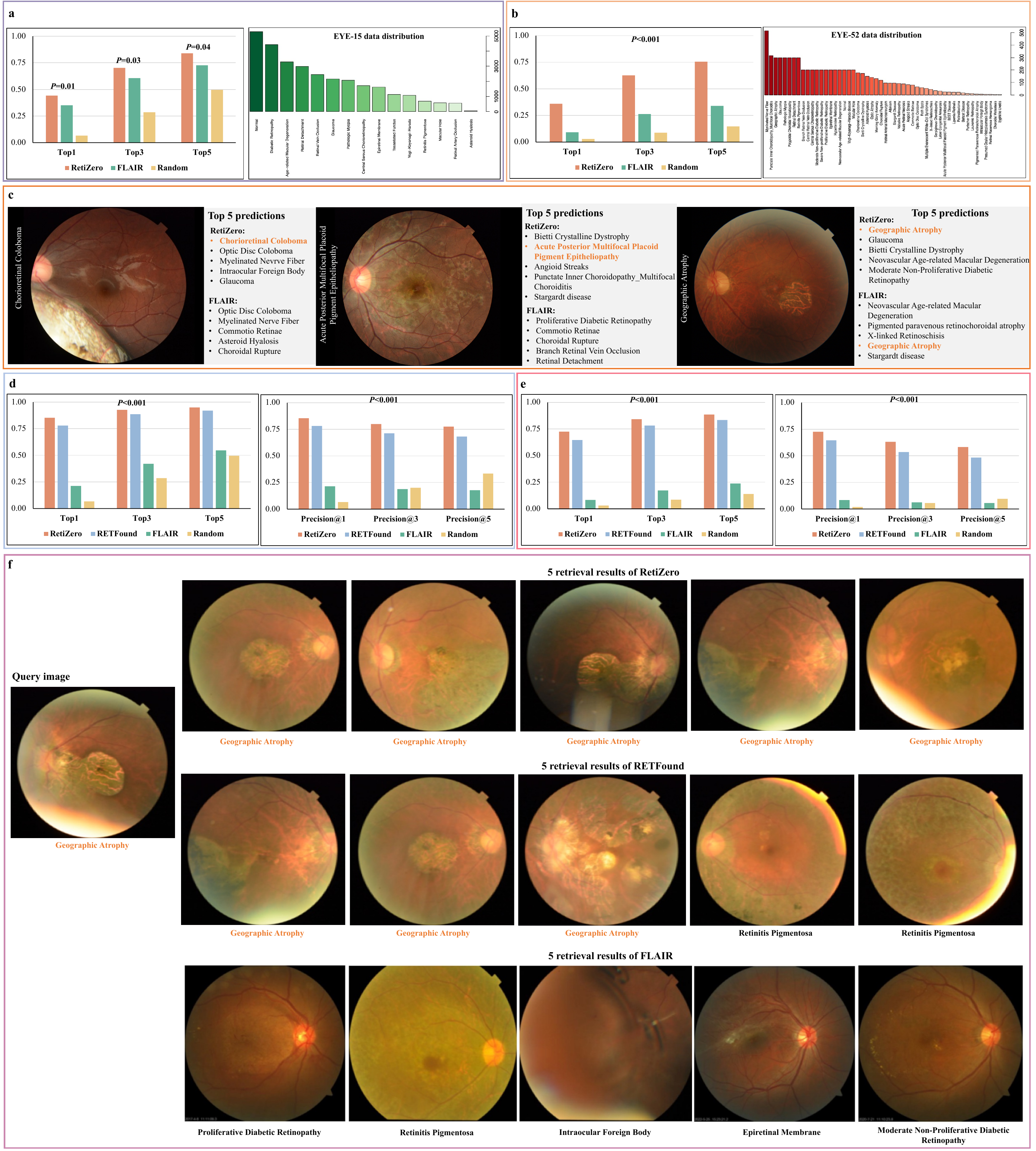}
 \end{center}
 \caption{\textbf{Overall Top-1, Top-3, and Top-5 scores for zero-shot based fundus disease recognition and Fundus disease identification by image-to-image retrieval.} a, The zero-shot performance on EYE-15 dataset, which contains 30,089 fundus images including 14 common fundus diseases and a normal condition. b, The zero-shot performance on the EYE-52 dataset, which contains 7,007 fundus images including 51 categories of fundus diseases and a normal condition. c, Zero-shot fundus diseases identification samples. d, Image-to-image retrieval performance on EYE-15 dataset. e, Image-to-image retrieval performance on the EYE-52 dataset. f, Image-to-image retrieval samples. All P values were calculated with the two-sided t-test and listed in the figure.}
\label{Zeroshot}
\end{figure}
\FloatBarrier

\begin{figure}[h]
 \begin{center}
  \includegraphics[width=0.85\textwidth, height=0.85\textheight, keepaspectratio]{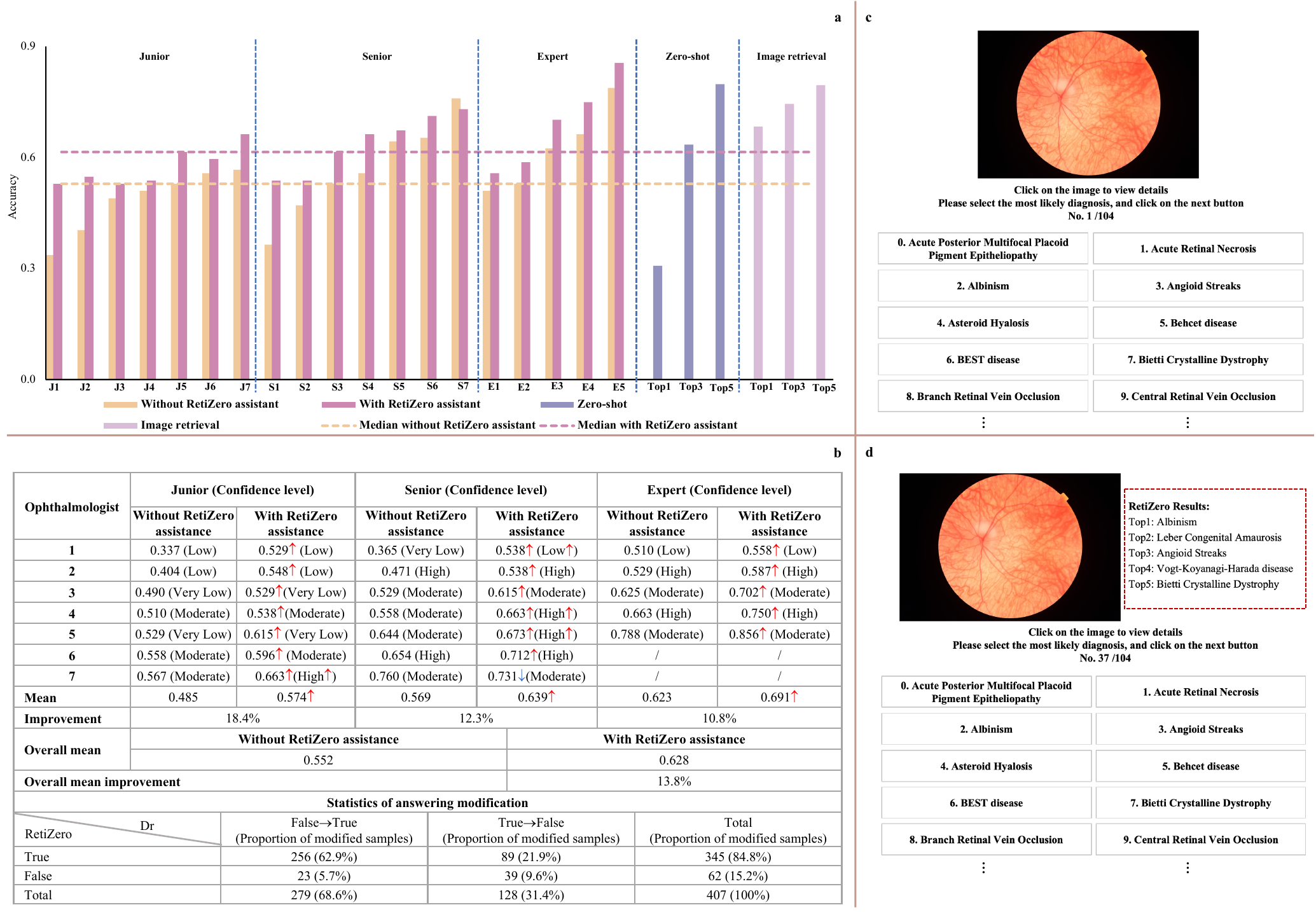}
 \end{center}
 \caption{\textbf{AI-assisted clinical diagnosis results.} a, Online fundus image reading system without RetiZero assistance. b, Online fundus image reading system with RetiZero assistance, c, Ophthalmologist diagnostic results, Top-1, Top-3, and Top-5 performance for zero-shot and image-to-image retrieval. d, Details for clinical evaluation.}
\label{ClinicalEvaluation}
\end{figure}
\FloatBarrier

\subsection{AI-assisted clinical diagnosis}
As illustrated in Task III in Fig. 1e, we first compared the performance of RetiZero with that of 19 ophthalmologists from Singapore, China and the U.S., and then examined whether RetiZero can assist in the clinical diagnosis made by ophthalmologists. In brief, for this clinical validation study, we created a subset of data comprising a total of 104 images by randomly selecting 2 samples per category from the EYE-52 set, and the ophthalmologists were asked to make a clinical diagnosis based on the retinal photographs without (Round 1, Fig. 3a) and with (Round 2, Fig. 3b, conducted one week after Round 1) the assistance of RetiZero.  The diagnostic accuracy of the 19 ophthalmologists ranged from 0.337 to 0.788, with the median of 0.582, while RetiZero's zero-shot Top-1, Top-3, and Top-5 accuracies are 0.308, 0.635, and 0.798, respectively (Fig. 3c). Therefore, RetiZero's zero-shot Top-3’s performance was comparable to that of the majority of ophthalmologists, while its Top-5’s performance surpassed that of all ophthalmologists. Furthermore, RetiZero’s performance the fundus disease identification by image-to-image retrieval for RetiZero achieved Top-1, Top-3, and Top-5 accuracies of 0.684, 0.745, and 0.796 surpassing that of the majority of ophthalmologists.

With the assistance of RetiZero, the performance of 18 out of the 19 ophthalmologists improved (Fig. 3c), out of 1,976 total responses (104 questions × 19 doctors), 1,569 responses (79.4\%) remained unchanged, while 407 responses (20.6\%) were modified in the second round. Of the 407 modified responses, 279 (68.6\%) were changed from incorrect to correct, demonstrating that most AI-assisted modifications helped the clinicians rectify previous misdiagnoses. Moreover, as shown in Fig. 3d, we further grouped the ophthalmologists by years of experience as junior ($\leq _latex$ 5 years, 7 doctors), senior (5-10 years, 7 doctors), and expert (>10 years, 5 doctors). Their average first-round accuracies were 48.5\%, 56.9\%, and 62.3\%, respectively, improving to 57.4\%, 63.9\%, and 69.0\% in the second round, with respective improvements of 18.4\% (P = 0.01), 12.3\% (P = 0.03), and 10.8\% (P < 0.001). This demonstrates that RetiZero-assisted diagnosis enhances accuracy across all experience levels, with junior clinicians benefiting the most. Additionally, the clinicians' confidence in their diagnoses increased from an average of 2.7 (low to moderate) in the first round to 3.0 (moderate, P = 0.04) in the second round, suggesting that RetiZero assistance not only improved accuracy but also boosted their confidence in making clinical diagnoses. Finally, there was a strong correlation between the top-ranking scores from RetiZero and the response modification scores of ophthalmologists (r = 0.614, P < 0.001, see Methods for details)), suggesting that the higher the correct result was ranked within the model's top 5 predictions, the greater the probability that the doctor would arrive at an accurate diagnosis. Moreover, to more clearly illustrate the effectiveness of RetiZero in assisted diagnostics, we further provided additional qualitative analysis in Extended Data Fig. 8-10.
\begin{figure}[h]
 \begin{center}
  \includegraphics[width=0.9\textwidth, height=0.9\textheight, keepaspectratio]{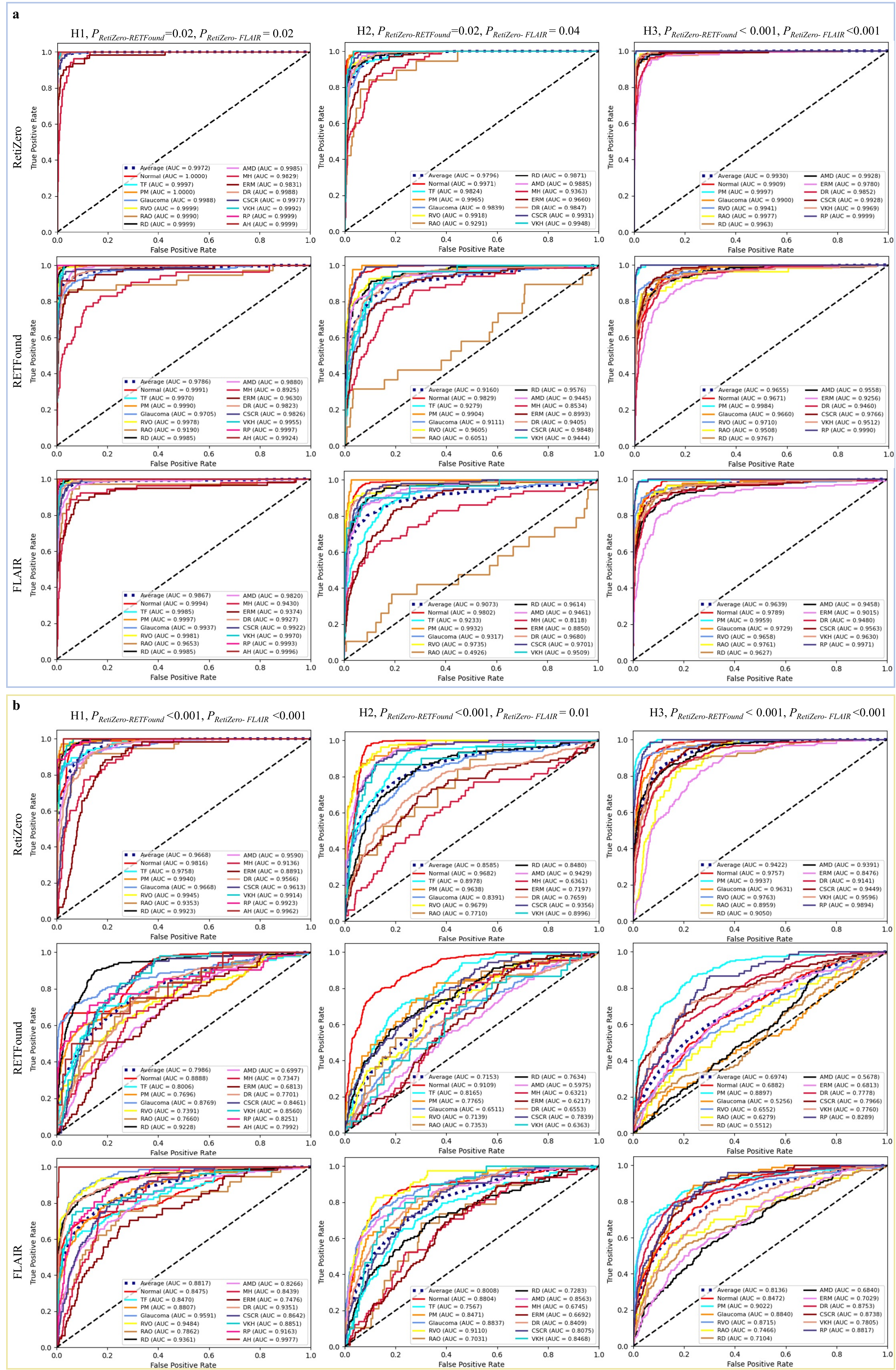}
 \end{center}
 \caption{\textbf{The receiver operating characteristic (ROC).} a, ROC curves for internal domain retinal disease identification. b, ROC curves for few-shot learning. The P values were calculated with the two-sided t-test.}
\label{Iden_Few}
\end{figure}
\FloatBarrier
\subsection{Internal domain fundus disease identification}
In Task IV (Fig. 1f), we collected three clinical datasets of retinal photographs, named as H1, H2, and H3, to validate the performance of RetiZero in internal domain fundus disease identification tasks. Extended Data Fig. 11 provides the data collection process and annotation details of three datasets. "Internal domain" means that we fine-tuned and validated the model separately within each of the three datasets. The details of three datasets are shown in Supplementary Tables 4 to 6. As shown in Fig. 4a, RetiZero achieved average AUCs of 0.997, 0.980, and 0.993 on the three datasets, respectively, each encompassing 15, 13, and 12 different categories of fundus diseases and/or normal condition, respectively. These results represent improvements of 1.9\% (P = 0.02), 6.9\% (P = 0.02) and 2.9\% (P < 0.001) compared to RETFound, and 1.1\% (P = 0.02), 8.0\% (P = 0.04) and 3.0\% (P < 0.001) compared to FLAIR. This is particularly apparent for certain fundus diseases with specific features, such as macular hole, epiretinal membrane, and retinal artery occlusion.

\subsection{Few-shot fine-tuning}
In task V (Fig. 1g), we fine-tuned the model using only five images from each fundus disease to evaluate RetiZero's performance with very limited training data. The data details are provided in Supplementary Tables 7 to 9. RetiZero achieved the highest AUC scores across the three datasets compared to RETFound and FLAIR (Fig. 4b). In the task of identifying 15, 13, and 12 types of fundus diseases in the H1, H2, and H3 dataset, RetiZero achieved AUC values of 0.967, 0.859, and 0.942 respectively, representing improvements of 7.2\% to 35.1\% over RETFound and FLAIR (all P < 0.001). These results indicated that even with limited annotated data samples, RetiZero can effectively learn the characteristic information of different fundus diseases in fundus images.

\subsection{Cross-domain fundus disease identification}
To assess the robustness of RetiZero in the task of cross-domain fundus disease identification (Task VI, Fig. 1h), we reorganized the three datasets of H1, H2, and H3 and only used the data with shared diseases categories across these datasets. Then we sequentially used one of the reorganized datasets, named rH1, rH2, and rH3, as internal testing sets and utilized the remaining two datasets as external testing sets to verify the robustness of different foundation models. The data information for different experimental strategies is presented in Supplementary Tables 10 to 12. As shown in Fig. 5, RetiZero achieved promising performance in all validation settings. In the internal testing set of the three datasets, RetiZero achieved AUC values of 0.998, 0.986, and 0.990, respectively, representing significant improvements of 3.1\%, 8.6\%, and 6.3\% over RETFound (all P < 0.001); and of 0.7\%, 4.4\%, and 5.0\% over FLAIR (all P < 0.001, Fig. 5a). In the external testing sets, the performance of RetiZero was similar to the internal testing set, with all AUCs $\geq$ 0.912, and significantly outperformed RETFound and FLAIR in all tasks (all P $\leq _latex$ 0.01, see in Fig. 5b and Fig. 5c). Additionally, RetiZero achieved outstanding performance in the identification of fundus diseases across most of the categories, especially in diseases with unique pathologic features such as epiretinal membrane, retinal artery occlusion, and central serous chorioretinopathy (Extended Data Fig. 12 to 14). We further conducted additional experiments to assess whether training with both domains rH1 and rH2, as opposed to training with only one of them, yields improved performance on domain rH3 (see Extended Data Fig. 15). The results indicate that incorporating both rH1 and rH2 during training leads to performance enhancements of 0.5\% and 0.6\% on rH3, with statistical significance (P = 0.029 and P = 0.008, respectively). These findings highlight the advantages of domain combination in achieving better generalization. Furthermore, we evaluated the inference times of various foundational models for diagnosing a single image (refer to Supplementary Table 13). RetiZero, RETFound, and FLAIR demonstrated inference times of 0.013, 0.012, and 0.017 seconds per image, respectively. Although RetiZero's inference time is marginally higher than RETFound's, it remains well within the acceptable range for real-time processing.

In addition, we rigorously tested our model's performance across multiple datasets representing diverse ethnic groups. This included two publicly available datasets (See Supplementary Table 14-15): the 2019 Sydney Innovation Challenge dataset (SIC, encompassing Caucasian and Indian populations, etc.,) and the SUSTech dataset (was collected from China) for Diabetic Retinopathy (DR) staging. Additionally, we utilized two in-house datasets for AMD and DR screening (See Supplementary Table 16-17): the Singapore Malay Eye Study (SiMES) dataset was collected from Malay individuals and the Singapore Indian Eye Study (SINDI) dataset was collected from Indian individuals. Across these diverse populations, RetiZero consistently outperformed RETFound and FLAIR (see Extended Data Figure 16). These comprehensive experimental results underscore RetiZero's robust performance and generalizability, affirming its efficacy in accurately diagnosing fundus diseases across varied demographic groups and enhancing its potential for widespread clinical application.

\begin{figure}[h]
 \begin{center}
  \includegraphics[width=\textwidth, height=1\textheight, keepaspectratio]{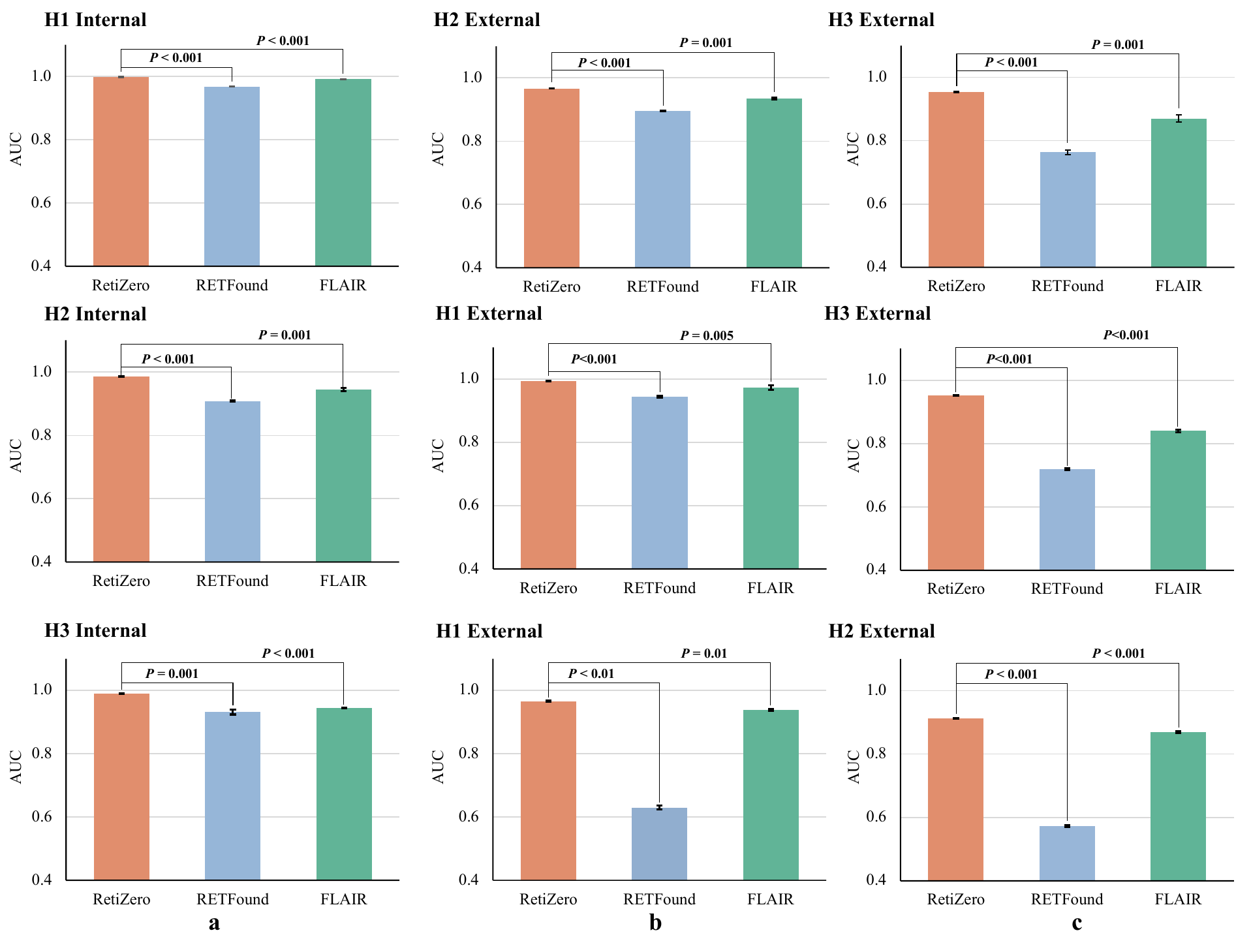}
 \end{center}
 \caption{\textbf{Cross-domain performance (AUC) of different foundation models for fundus disease screening.} column a, Internal evaluation: Different foundation models were adapted to each dataset by fine-tuning and internally evaluated on hold-out testing data. Columns b-c, performance on external validation sets: The three foundation models were tested on the other two external validation datasets. The disease categories and dataset strategy information are listed in Supplementary Tables 10-12. The error bars show 95\% CI and the bar centre represents the mean value of the AUC. P value was calculated with the two-sided t-test and listed in the figure.}
\label{Gen}
\end{figure}
\FloatBarrier

\section{Discussion}
In this study, we trained a vision-language-foundation model, RetiZero, using a vast amount of image-text pairs. Comprehensive experimental results demonstrated that RetiZero has strong capability in representing fundus disease features across a wide range of downstream tasks of fundus disease identification, including zero-shot recognition, image-to-image retrieval, AI-assisted clinical diagnosis, internal domain and cross domain classification, and few-shot fine-tuning. The performance of RetiZero is superior to two state-of-the-art ophthalmic LFMs, RETFound~\cite{RETFound} and FLAIR~\cite{FLAIR}. These results collectively demonstrated the superior generalizable and robust performance of RetiZero in both common and rare fundus disease identification.

The superiority of RetiZero over RETFound~\cite{RETFound} and FLAIR~\cite{FLAIR} can be attributed to its unique design and diverse data used for pre-training. Specifically, although the RETFound model~\cite{RETFound}, pre-trained on a large number of fundus images using the MAE architecture, can enhance the performance of various downstream tasks, it includes limited number of fundus disease categories, particularly rare fundus diseases. In addition, it lacks the incorporation of textual information, resulting in inadequate characterization of image feature attributes. This makes it unsuitable for text prompt-based zero-shot fundus disease detection tasks, thereby limiting its application in clinical practice, particularly for the identification of rare fundus diseases. In contrast, FLAIR~\cite{FLAIR}, based on the CLIP architecture, incorporates textual description information during network training to enhance the representation of image feature attributes. However, it was pre-trained on a very limited dataset of fundus disease knowledge, leading to poor performance in zero-shot recognition tasks for rare fundus diseases. Furthermore, FLAIR~\cite{FLAIR} and other standard CLIP-based models~\cite{huang2023visual,radford2021learning} lack guidance for learning information such as lesion contours and topological structures in images, resulting in low performance in fundus disease identification through image-to-image retrieval. In contrast to these existing foundation modes, RetiZero leverages the synergy between MAE-based self-supervision and contrastive alignment with a broad set of disease descriptions to enhance its disease recognition capabilities. By combining the detailed feature extraction of MAE, which excels in capturing visual details through unsupervised learning from large-scale image datasets, with the semantic richness of contrastive text-image alignment, RetiZero gains a stronger implicit representation of lesions. This approach not only retains the granularity of local visual features critical for identifying subtle disease markers but also aligns these features with a diverse range of textual disease descriptors. Consequently, RetiZero achieves superior performance by effectively bridging the gap between high-level semantic understanding and low-level visual details, making it adept at recognizing a wide array of fundus conditions, both common and rare.

Developing artificial intelligence algorithms to assist in the clinical diagnosis of rare fundus diseases has been challenging. Although image classification with few-shot learning has shown promising results~\cite{quellec2020automatic,gao2023discriminative}, it is challenging to collect enough samples for training for extremely rare diseases, especially considering there are more than 400 fundus diseases. Zero-shot learning would be a better approach for this scenario. In previous literatures, there are only few reports investigating zero-shot learning for fundus images, which only focus on diabetic retinopathy or image quality assessment, and struggle with rare conditions~\cite{yi2023label,mahapatra2022self}. The image-to-image retrieval approach is another possible solution. Current literature only investigated diabetic retinopathy, not rare diseases, and the results showed limited performance, possibly due to the lack of integration of textual information crucial for complex disease identification~\cite{fang2021deep,fang2023deep,chandakkar2012machine,quellec2011automated}. To our knowledge, our study is the first one to use zero-shot learning and image-to-image retrieval to cope with the challenge of rare disease diagnosis. RetiZero leverages both the MAE and CLIP architectures to enhance feature representation from diverse datasets, achieving superior performance in zero-shot fundus disease identification and image-to-image retrieval tasks. By integrating the MAE and CLIP architectures, RetiZero bridges the semantic gap, incorporates knowledge of over 400 types of fundus diseases, and has been validated on a clinical dataset encompassing more than 50 types.

Diagnosis of rare diseases is also very challenging for clinicians. Our results show that the accuracy of junior, senior, and expert ophthalmologists was 0.485, 0.569, and 0.623, respectively, in classifying the testing set with 104 retinal photos of 52 categories. In a clinical scenario, clinicians may also list several diseases for differential diagnosis, and then integrate other information from history, ocular examination, and other investigations for differential diagnosis. Our study mimics this approach. Although the zero-shot Top-1 accuracy (0.360) of RetiZero was lower than that of the ophthalmologists, the zero-shot Top-3 (0.626) and Top-5 (0.756) accuracies of RetiZero were comparable to and exceeded those of most ophthalmologists, respectively. This approach serves as a valuable reference for ophthalmologist in differential diagnosis. Our results also showed that using the zero-shot Top-5 results from RetiZero significantly enhanced the ophthalmologists’ diagnostic accuracy, with expert ophthalmologists improving by 10.8\% and junior ophthalmologists by 18.4\%. Moreover, the correlation between RetiZero's reference predictions and the accuracy improvements observed in ophthalmologists underscores its potential to assist in differential diagnoses. In the exploration of AI in the medical field, initial studies primarily focused on comparing the performance of AI algorithms to physicians (AI-physician comparison) to the performance of AI algorithms~\cite{dong2022artificial,ruamviboonsuk2022real,li2024performance}. However, in real-world implementation, AI algorithms can make errors, posing potential risks to patient outcomes.  To mitigate the risks, therefore, AI-assisted diagnosis or AI-physician collaboration have been proposed and investigated. Various approaches have been reported, such as using AI algorithms to suggest the most likely diagnosis, which physicians then confirm~\cite{lin2021application}. Additionally, AI models have been shown to improve the diagnostic performance of junior doctors~\cite{li2022development}. However, these studies mostly focus on common diseases with their AI models typically provide only one most probable disease diagnosis. In contrast, our RetiZero offers the top five results using the Zero-shot approach, significantly enhancing the diagnostic accuracy of ophthalmologists with various levels of clinical experiences. Beyond its robust zero-shot recognition capabilities, RetiZero’s alignment of structured medical text with fundus imagery opens new avenues for practical clinical integration and cross-institutional collaboration. For example, busy ophthalmologists can rapidly screen rare pathologies—such as Bietti crystalline dystrophy or chorioretinal coloboma—without pre-training on large labelled sets, facilitating timely triage or referral in resource-limited clinics. Its image-to-image retrieval function helps confirm diagnoses by surfacing visually similar cases from vast, heterogeneous archives, which is particularly beneficial when local datasets are small or unbalanced. Moreover, RetiZero’s top-5 differential diagnosis output can enhance the confidence and accuracy of both junior and expert clinicians, making it especially attractive for teleophthalmology networks or multicenter screening programs.

We also recognized limitations and areas for improvements in the current work. Although our datasets include knowledge of over 400 types of fundus diseases, the imbalance across different categories may limit RetiZero's performance in downstream tasks. To address this, we plan to further enrich the dataset with a more balanced representation of various fundus diseases, particularly rare ones. Potentially, addressing data imbalance between rare and common ophthalmic pathologies could involve advanced synthetic data generation and tailored augmentation pipelines aimed specifically at underrepresented classes. Generative Adversarial Networks (GANs) or diffusion models can create realistic fundus images that mirror rare disease appearances, thereby expanding the sample diversity and allowing the network to learn clinically important features that are otherwise scarce~\cite{gui2021review,rombach2022high}. Beyond straightforward geometric transformations (e.g., rotations, flips), lesion-centered augmentations—like elastic deformations or localized color perturbations—further enrich minor-class variance without losing critical pathological signatures~\cite{shorten2019survey}. These methods systematically enhance the representation of difficult or rare pathologies, ultimately bolstering model robustness in the face of severe data imbalance. In addition, while RetiZero has shown promising performance across multiple tasks and datasets, specialized models optimized for specific tasks may outperform more generalized models. Therefore, we will continue to explore targeted improvements to enhance RetiZero’s performance for specific applications.

In conclusion, the proposed feature-calibrated fundus vision-language foundation model, RetiZero, which incorporates knowledge of over 400 fundus diseases, effectively capture the rich contextual feature in fundus images and learn the alignment between retinal image features and disease-related textual information. RetiZero demonstrated superior performance in feature representation and generalizability across different fundus disease recognition tasks, tested at multiple eye clinics using different fundus cameras, under different degrees of domain drift, and with very limited training samples. Notably, the excellent performance of RetiZero’s exceptional performance in zero-shot fundus disease identification and image-to-image retrieval-based is of significant value for screening fundus diseases in clinical practice, particularly rare fundus conditions. Furthermore, comprehensive AI-assisted clinical diagnosis results further confirm that RetiZero can enhance ophthalmologists' diagnostic accuracy and confidence, offering particular benefit to less experienced clinicians.

 \section{Methods}
\subsection{Dataset}
\subsubsection{Data for pretraining: }
RetiZero uniquely integrates an MAE-based backbone with a CLIP-style contrastive framework and uncertainty-based feature calibration to achieve robust image-text alignment across more than 400 fundus disease categories. RETFound was pre-trained using a Masked Autoencoder (MAE) approach on over 900,000 fundus images. This large-scale pretraining allows RETFound to develop a deep understanding of fundus image features. Therefore, we utilized RETFound as the backbone of our image encoder. This approach allows us to enhance RetiZero's ability to effectively handle complex fundus image data by building on a robust, pre-established feature extractor. We pre-trained RetiZero using a dataset we collected, comprising of 341,896 image-text pairs that cover over 400 fundus diseases. As shown in Fig. 1a, the image-text pretraining data mainly consists of three parts: publicly available dataset with category information, data from the ophthalmic literature with disease-related keywords descriptions, and image-text pairs from online resources (Supplementary Table 13). Specifically, we collected a total of 303,129 fundus images from 29 publicly available datasets, encompassing over 100 distinct categories of fundus diseases. To ensure consistency in the presentation of image descriptions, we converted the fundus images with classification labels into a uniform textual description format (Fig.1b). In addition, we engaged 12 ophthalmologists to compile and curate a comprehensive dataset. They reviewed 180 ophthalmic literature sources and online resources, identifying and documenting 23,228 fundus images with corresponding disease-related textual descriptions from the literature, as well as 15,544 fundus images from online resources. A team of ten ophthalmologists was assigned to this task. Their responsibilities included browsing various sources to locate fundus photographs, capturing screenshots of these images, and recording the source (literature/website), page number (book/page), and extracting the fundus description that uniquely corresponded to each respective fundus image. Our selection process was rigorous, including only those textual descriptions that precisely matched their corresponding color fundus photographs. Any information unrelated to the image, such as titles, background details, or descriptions of other imaging modalities or outside the field of view, was deliberately excluded. Furthermore, these ten ophthalmologists also assigned a diagnostic label for each image simultaneously. The final dataset underwent a thorough curation and validation process by two experienced ophthalmologists. Their responsibilities included unifying the labels to reduce label noise and excluding non-standard images, such as blurry, over- or under-exposed, montage, local view, monochrome images, and those annotated with additional graphics, thereby reducing image noise. Finally, all curated disease and lesion labels were converted into the standardized format "a fundus image of [disease/lesion labels]" and input into the text encoder model.

As shown in Supplementary Table 14, these images cover 414 ophthalmic labels, encompassing nearly all known fundus diseases to date. Meanwhile, Supplementary Table 15 provides further details about 180 ophthalmic literature sources. Furthermore, we collected 28,800 fundus images with relevant descriptions from the online resources. The 12 ophthalmologists manually cleaned and organized 15,544 images along with their corresponding disease-related textual descriptions. We pre-trained RetiZero using the PyTorch framework on an Nvidia Geforce DGX A100 GPU (80G), with a batch size of 128 and the Adam optimizer. The data collection process for RetiZero pretraining is illustrated in Fig. 1a. 

The use of images from literature and certain websites in this research should constitute fair use and should accordingly not constitute copyright infringement. Our pre-trained model does not generate any visible data similar to the images from these resources, nor is any data from the literature and websites reused in other downstream tasks. Importantly, the pre-trained model is sorely for academic research and is not intended for commercial purposes. In summary, the dataset for pretraining RetiZero covers nearly all known fundus diseases, integrating very comprehensive ophthalmic knowledge.

\subsubsection{Data for internal domain fundus disease identification: }
To evaluate the performance of RetiZero in the task of fundus disease identification, we curated three retinal photos datasets across multiple hospitals: 1) healthcare dataset 1 (H1) where retinal photos were acquired using  the Topcon TRC-NW8 and Zeiss VISUCAM-200 devices at the Joint Shantou International Eye Center (JSIEC); 2) healthcare dataset 2 (H2) where retinal photos was acquired by the Topcon TRC-NW8 and Zeiss VISUCAM-200 devices from four hospitals, including Longchuan People's Hospital, Heyuan; Puning People's Hospital, Jieyang;  Wuping Hospital, Longyan; and Pengpai Memorial Hospital, Shanwei; 3) healthcare  dataset 3 (H3) was acquired by the Topcon DRI OCT Triton device of the (JSIEC). The H1 and H2 datasets exhibit domain differences due to varying clinic sources, whereas the H3 dataset differs from both H1 and H2 due to the use of distinct fundus images acquisition devices. Clinical assessment and labeling procedure are summarized in Extended Data Fig. 11. This study was approved by the JSIEC Institutional Review Board and adhered to the principles of the Declaration of Helsinki. All fundus images used in this study were made de-identified before computational analysis and model development.
The H1 dataset consisted of 11,414 fundus images, covering 15 categories of fundus diseases and normal condition. We further divided the H1 dataset into training (no. of images = 6,942), validation (n = 2,284), and testing (n = 2,288) for model fine-tuning, selection, and performance verification, respectively (Supplementary Table 4). The H2 dataset consisted of 7,812 fundus images, representing 12 types of fundus diseases and normal condition (Supplementary Table 5). To validate the performance of fine-tuning RetiZero for fundus disease identification on the H2 dataset, we partitioned the H2 dataset into training (n = 4,682), validation (n = 1,561), and testing (n =1,569) set, respectively, for model fine-tuning, selection, and performance evaluation, respectively. The H3 dataset consisted of 10,863 fundus images across 12 disease categories. It was divided into training (n = 6,511), validation (n = 2,174), and testing (n = 2,178) sets for model fine-tuning, selection, and performance evaluation, respectively (Supplementary Table 7). In this task, RetiZero was fine-tuned for internal domain fundus disease identification using PyTorch on an Nvidia Geforce 3090 GPU (24G). Adam optimizer and cross-entropy loss function were adopted to guide the model fine-tuning. The total iteration epoch and batch size were set to 100 and 64, respectively.

\subsubsection{Data for few-shot fine-tuning: }
To assess RetiZero’s performance in the few-shot fine-tuning, we reorganized the H1, H2, and H3 datasets. Specifically, we randomly selected 5 samples from each disease category of H1, H2, and H3 training set for few-shot fine-tuning, while retaining the validation and testing datasets for model selection and performance evaluation. Detailed information on category and data distribution is provided in Supplementary Tables 7-9. In this task, RetiZero was fine-tuned using PyTorch on an Nvidia Geforce 3090 GPU (24G). The Adam optimizer and cross-entropy loss function were employed for model optimization, with a total of 1,000 epochs and a batch size of 32.

\subsubsection{Data for cross-domain fundus disease identification: }
To evaluate the generality and robustness of RetiZero in cross-domain fundus disease identification, we again reorganized the H1, H2, and H3 datasets. This process identified 11 overlapping diseases categories across the three datasets, which were subsequently renamed as rH1 (no. of images = 10,304), rH2 (n = 6,829), and rH3 (n = 10,485). Details are provided in Supplementary Tables 10 to 12. We then performed three experimental settings. Specifically, rH1, rH2, and rH3 were sequentially used as internal datasets for model fine-tuning, selection, and internal testing, while the remaining two datasets served as external testing sets to assess. In this task, RetiZero was fine-tuned using PyTorch on an Nvidia Geforce 3090 GPU (24G). The Adam optimizer and cross-entropy loss function were employed, with the model trained over 100 epochs and a batch size of 64.

\subsubsection{Data for the tasks of zero-shot and image-to-image retrieval:}
We combined the three datasets, H1, H2, and H3 from different hospitals into a dataset named EYE-15, containing 30,089 fundus images that included 14 common fundus diseases and 1 normal category. This dataset was used to validate RetiZero's performance in screening common fundus diseases in zero-shot and image-to-image retrieval tasks. The no. of images in each disease category in EYE-15 was provided in Supplementary Table 1. Moreover, we further collect additional 7,007 fundus images acquired by different fundus cameras (such as Topcon TRC-NW8, Zeiss VISUCAM-200, and Topcon DRI OCT Triton) from JSIEC and the other four hospitals (EYE-52 dataset), comprising of 51 fundus diseases and 1 normal condition, to validate the performance of zero-shot fundus disease recognition and fundus disease identification by image-to-image retrieval in a more challenging setting. As shown in Supplementary Table 2, EYE-52 included a wide range of clinically rare fundus diseases, such as albinism, Bietti crystalline dystrophy, choroidal coloboma, and choroidal neoplasm. We adopted Top-1, Top-3, and Top- 5 accuracy to evaluate the performance of RetiZero in both tasks of zero-shot fundus disease recognition and fundus disease identification by image-to-image retrieval. Top-K accuracy is a metric used to evaluate the performance of a classification model by determining whether the correct label for a given input appears within the Top-K predictions made by the model, as follows:
\begin{equation}
    Top-K=\frac{\sum_{i=1}^{N} 1\left\{ y_{i}\in Top\left( \hat{y}_{i} \right) \right\}}{TotalnumberofinputsN},
\end{equation}
where $1\left\{ \bullet \right\}$ is the indicator function that evaluates to 1 if the ground truth label is in the topk predictions, and 0 otherwise. Extended Data Fig. 11 provides the process of the collection for the EYE-15 and EYE-52 datasets. In this study, we also adopted Precsion@1, Precesion@3, and Precsion@5 as the metrics to evaluate the performance of different foundation models in the task of fundus disease retrieval. Precision@N is a metric used to evaluate the performance of information retrieval systems and ranking algorithms. It is specifically used to measure the precision of the Top- N results returned by a system. Here is the formula and its explanation:
\begin{equation}
    Precision@N=\frac{|R_{all}\cap R_{Re}@N|}{N},
\end{equation}
where $R_{all}$ is the set of all relevant samples for the given query, $R_{Re}$ represents the set of the Top-N samples retrieved by the system in response to the query. $|R_{all}\cap R_{Re}@N|$ denotes the number of relevant samples in the Top-N retrieved documents, that is the count of samples that are both relevant and retrieved within the Top-N results, while N is the number of Top-samples considered for the calculation.

\subsection{Framework of RetiZero:} Fig. 1b provides an overview of the RetiZero framework. RetiZero integrates the advantages of MAE self-supervised learning and CLIP contrastive learning architectures. Specifically, the model was built upon the MAE-based pre-trained backbone network RETFound16, whose weights were frozen to preserve the model's representation capability for complex semantic information such as lesion contours and topological structures in retinal images. Meanwhile, we introduced low-rank learnable factors into the pre-trained RETFound and leveraged the CLIP architecture to learn image-text knowledge, aiming to enhance the model's understanding of image-text correlations and improve its feature representation capabilities. Furthermore, we incorporated an uncertainty vision-language feature calibration method based on Dirichlet reparameterization into the contrastive vision-language pretraining framework. This further refined visual-language features in the high-dimensional embedding space, thereby enhancing the model's ability to represent complex features in fundus images. As a result, RetiZero was developed, integrating the advantages of both MAE and CLIP architectures to provide robust feature support for downstream tasks. The components of RetiZero are discussed in details in the following sections.

\subsubsection{Image Encoder:}
As shown in \textbf{Figure~\ref{Overview}}, the image encoder consists of MAE-based SSL pre-trained backbone and low-rank learnable factors. MAE is a widely used self-supervised learning approach that employs a simple autoencoder approach to reconstruct the original signal based on partial observations. MAE-based SSL pretraining can guide the network to focus on the rich structural information and contextual features in the images. Therefore, RETFound~\cite{RETFound}, pre-trained on over 900,000 fundus images, is adopted as our MAE-based pre-trained backbone. Low-rank learnable factors (LoRA) are a parameter-efficient transfer learning method based on reparameterization~\cite{hu2021lora}, which utilizes low-rank representations to minimize the number of trainable parameters. It enables a pre-trained large foundation model to incorporate new knowledge into new target tasks, demonstrating robust and state-of-the-art (SOTA) performance in various parameter-efficient transfer learning tasks. Therefore, we utilize low-rank learnable factors to introduce retinal feature description information into the image encoder of RetiZero, enhancing its capacity to represent feature attributes of retinal images. Specifically, given the input token sequence $F_{in}\in R^{B\times N\times C_{in}}$ and the output token sequence $F_{out}\in R^{B\times N\times C_{out}}$ obtained by the projection layer $W\in R^{C_{out}\times C_{in}}$, LoRA assumes that updates to $W$ should be gradual and stable. Therefore, we apply low-rank approximations to delineate this gradual update. First, freeze the transformer layer to keep W fixed while adding a bypass to complete the low-rank approximation. And, the bypass consists of two linear mapping layers, $A\in R^{r\times C_{in}}$ and $B\in R^{C_{out}\times r}$, where $r\ll \left\{ C_{in},C_{out}\right\}$. Thus, the processing of the update layer $\hat{W}$ can be described as:
\begin{equation}
F_{out}=\hat{W} F_{in},
\end{equation}

\begin{equation}
\hat{W} =W+\nabla W=W+BA.
\end{equation}
The multi-head self-attention mechanism identifies the regions of focus using feature-relevant intensities, and previous works showed that focusing adaptations on Q and V matrices in the transformer's attention mechanism can effectively capture task-specific nuances without the need for extensive retraining or increasing model complexity~\cite{jie2023fact,chen2024ma}. Therefore, we apply low-rank approximation to the projection layers of the query and value to influence the attention scores.:

\begin{equation}
Att\left( Q,K,V\right)  =Softmax\left( \frac{QK^{T}}{\sqrt{C_{out}} } +B\right)  V,
\end{equation}

\begin{equation}
Q=\hat{W}_{q} F=W_{q}+B_{q}A_{q}F,
\end{equation}

\begin{equation}
K=W_{k}F,
\end{equation}

\begin{equation}
V=\hat{W}_{v} F=W_{v}F+B_{v}A_{v}F,
\end{equation}
where $W_{q}$, $W_{k}$, and $W_{v}$ are frozen projection layers of RETFound, while $A_{q}$, $B_{q}$, $A_{v}$, and $B_{v}$ are trainable LORA factors. 

\subsubsection{Text Encoder: }
Descriptions of fundus images are typically more challenging than those of natural images, as they often contain numerous specialized clinical terms, sometimes even comprising of multiple lesion signs or sentences. Therefore, in this paper, we utilize the BioClinicalBERT~\cite{alsentzer2019publicly} model as the text encoder for extracting textual features. We use its pre-trained weights on medical texts from the MIMIC III dataset for weight initialization and train it based on a vision-language contrastive learning strategy to extract ophthalmic textual features.

\subsection{Uncertainty-based feature calibration}
Our dataset is collected from various global sources, including public databases, ophthalmic literature, and online resources. Inevitably, it includes complex data distributions such as differences in image resolution, incomplete textual descriptions, and low-quality image-text pairs. Therefore, focusing on these issues, we further introduced an uncertainty vision-language feature calibration method based on Dirichlet reparameterization~\cite{han2020trusted,ng2011dirichlet} into the contrastive vision-language pretraining framework, to further calibrate visual-language features in the high-dimensional embedding space for enhancing the robustness of the model to represent complex features in fundus images. Specifically, as shown in Figure 1, RetiZero's pretraining consists of a fundus image encoder and a text encoder. The linear layer serves as a projection head for both the image encoder and the text encoder, mapping the acquired features to a 512-dimensional embedding feature space. Let assume $\phi =\left\{ \phi_{E} ,\phi_{H} \right\}$ denotes image encoder ($\phi_{E}$) and corresponding projection head ($\phi_{H}$). Given a fundus image $X_{i}$, the image encoder is adopted to obtain feature representation of $F_{Img}=\phi_{E} \left( X_{i}\right)$. Meanwhile, $\psi =\left\{\psi_{E},\psi_{H} \right\}$ is used to represent text encoder ($\psi_{E}$) and corresponding projection head ($\psi_{H}$). The text encoder ($\psi_{E}$) is adopted to extract feature embedding $F_{T}=\psi_{E} (X_{T})$ from text input $X_{T}$.Then, image projection head ($\phi_{H}$) and text projection head ($\psi_{H}$) are utilized to map the independent modality representations into a joint unit hyper-sphere space: $I=\frac{\phi_{H} \left( F_{Img}\right)  }{\| \phi_{H} \left( F_{Img}\right)  \| }$ and $T=\frac{\psi_{H} \left( F_{T}\right)  }{\parallel \psi_{H} \left( F_{T}\right)  \parallel }$, respectively. The similarity between the input image ($X_{i}$) and input text ($X_{T}$) are evaluated by the cosine similarity based on the normalized features: $I^{Tr}T_{T}$, where $Tr$ represents the transpose operator. With obtained similarity metrics, the optimization goal of the contrastive-based learning pre-training approach is to minimize the distance of features between paired images and text descriptions while maximizing the distance between features of unpaired samples. Specifically, assuming that a batch contains $N$ samples, $I_{i}\in \left\{ I_{1},I_{2},...,I_{N}\right\}$ and $T_{i}\in \left\{ T_{1},T_{2},...,T_{N}\right\}$, represent image feature vector and text feature vector of each sample, while and $G=\left\{ 0,1,...,N-1\right\}$ is the corresponding category label, respectively. To guide model optimization, we use the following loss function. 

\begin{equation}
    L_{Con}=L_{Em}+L_{Dl},
\end{equation}

\begin{equation}
L_{Em}=\frac{1}{2} \left( \sum\nolimits^{N}_{i=1} -log\left( \frac{exp\left( I^{Tr}_{i}T_{i}\right)  }{\sum\nolimits^{N}_{k-1} exp\left( I^{Tr}_{i}T_{k}\right)  } \right)  +\sum\nolimits^{N}_{i=1} -log\left( \frac{exp\left( T^{Tr}_{i}I_{i}\right)  }{\sum\nolimits^{N}_{k-1} exp\left( T^{Tr}_{i}I_{k}\right)  } \right)  \right),
\label{lem}
\end{equation}
$L_{Dl}$ is a loss function based on feature vectors which are reparametrized from similarity measures using the Dirichlet distribution. The specific implementation is as follows:

\textbf{Step (1):} Obtaining the evidence feature $E_{I2T}$ and $E_{T2I}$ by applying the Softplus activation function to similarity metrics between image and text feature embedding to ensure the feature values are larger than 0:
\begin{equation}
    E_{I2T}=Softplus(I^{Tr}T), and E_{T2I}=Softplus(T^{Tr}I),
\end{equation}
where I2T and T2I indicate image-to-text and text-to-image contrastive direction.

\textbf{Step (2):} Parameterizing $E_{I2T}$ and $E_{T2I}$ to Dirichlet distribution, as:
\begin{equation}
    \bm{\alpha_{I2T,k}} =E_{I2T,k}+1,\  i.e.,\alpha_{I2T,k} =e_{I2T,k}+1,e_{I2T,k}=\left\{ Softmax\left( I^{Tr}_{k}T_{1}\right)  ,...,Softmax\left( I^{Tr}_{k}T_{N}\right)  \right\},\  
\end{equation}
\begin{equation}
    \bm{\alpha_{T2I,k}} =E_{T2I,k}+1,\  i.e.,\  \alpha_{T2I,k} =e_{T2I,k}+1,e_{T2I,k}=\left\{ Softmax\left( T^{Tr}_{k}I_{1}\right)  ,...,Softmax\left( T^{Tr}_{k}I_{N}\right)  \right\},
\end{equation}
where $\bm{\alpha_{I2T,k}}$, $\bm{\alpha_{T2I,k}}$, $e_{I2T,k}$, and $e_{T2I,k}$ are the $k$-th contrastive similarity Dirichlet distribution parameters and evidence for the image-text contrastive similarity of the $k$-th sample in a batch of $N$ samples.

\textbf{Step (3):} Calculating the belief masses and corresponding uncertainty score as:
\begin{equation}
    b_{I2T,k}=\frac{e_{I2T,k}}{S_{I2T}} =\frac{\alpha_{I2T,k} -1}{S_{I2T}} ,\  u_{I2T}=\frac{N}{S_{I2T}},
\label{BleiefI2T}
\end{equation}

\begin{equation}
    b_{T2I,k}=\frac{e_{T2I,k}}{S_{T2I}} =\frac{\alpha_{T2I,k} -1}{S_{T2I}} ,\  u_{T2I}=\frac{N}{S_{T2I}},
\label{BleiefT2I}
\end{equation}
where $S_{I2T}=\sum\nolimits^{N}_{k=1} \left( e_{I2T,k}+1\right)  =\sum\nolimits^{N}_{k=1} \alpha_{I2T,k}$ and $S_{T2I}=\sum\nolimits^{N}_{k=1} \left( e_{T2I,k}+1\right)  =\sum\nolimits^{N}_{k=1} \alpha_{T2I,k}$ are the Dirichlet intensities of image-to-text and text-to-image, respectively, used to constrain $1=\sum\nolimits^{N}_{k=1} b_{I2T,k}+u_{I2T}$ and $1=\sum\nolimits^{N}_{k=1} b_{T2I,k}+u_{T2I}$
It can be seen from Eq.~\ref{BleiefI2T}, and Eq.~\ref{BleiefT2I} the probability assigned to $k$-th sample is proportional to the observed similarity evidence for sample $k$. Conversely, if less total similarity evidence was obtained, the greater the total uncertainty. 

In this study, we associate the Dirichlet distribution with the distribution of feature similarity between images and text descriptions, thereby obtaining belief masses and corresponding overall uncertainty score for the similarity of images and text description for each sample of a batch, based on the evidence collected from the feature similarity matrix. Therefore, we could work out the Dirichlet distribution parameter of $\bm{\alpha_{I2T}} =\left[ {}\alpha_{I2T,1} ,...,\alpha_{I2T,N} \right]$ and $\alpha_{T2I} =\left[ {}\alpha_{T2I,1} ,...,\alpha_{T2I,N} \right]$ for image-to-text, and text-to-image, while obtaining the multinomial opinions $D\left( p_{I2T,i}|\alpha_{I2T,i} \right)$ and $D\left( p_{T2I,i}|\alpha_{T2I,i} \right)$, where $p_{I2T,i}$ and $p_{T2I,i}$ were the sample assignment probabilities on a simplex. Therefore, the loss function for the reparameterized similarity matrix as follows:
\begin{equation}
    L_{Dl}=L^{I2T}_{Dl}+L^{T2I}_{Dl},
    \label{Ldl}
\end{equation}
where,
\begin{equation}
    L^{I2T}_{Dl}=L^{I2T}_{Dl-CE}+\lambda \ast L_{KL}, 
\end{equation}
\begin{equation}
    L^{T2I}_{Dl}=L^{T2I}_{Dl-CE}+\lambda \ast L_{KL},
\end{equation}
where $L_{Dl-CE}$ ($L^{I2T}_{Dl-CE}$ and $L^{T2I}_{Dl-CE}$) was used to ensure that the correct prediction for the sample with highest similarity between image and text yielded more evidence than other samples, while $L_{KL}$ was used to ensure that incorrect predictions would yield less evidence, and $\lambda$ was the balance factor that was gradually increased so as to prevent the model from paying too much attention to the $KL$ divergence in the initial stage of training, which might result in a lack of good exploration of the parameter space and cause the network to output a flat uniform distribution.
\begin{equation}
    L_{Dl-CE}=\int \left[ \sum\nolimits^{N}_{k=1} -y_{k}log\left( p_{k}\right)  \right]  \frac{1}{\beta \left( \alpha_{i} \right)  } \prod\nolimits^{N}_{k=1} p^{\alpha_{k} -1}_{k}dp_{k}=\sum\nolimits^{N}_{k=1} y_{k}\left( \psi \left( S_{k}\right)  -\psi \left( \alpha_{k} \right)  \right),
\end{equation}
where $\psi \left(\right)$ was the digamma function, while $\beta \left(\right)$ is the multinomial beta function for the concentration parameter $\bm{\alpha}$.
\begin{equation}
    L_{KL}=log\left( \frac{\Gamma \left( \sum\nolimits^{N}_{k=1} \hat{\alpha }_{k} \right)  }{\Gamma \left( N\right)  \prod\nolimits^{N}_{k=1} \Gamma \left( \sum\nolimits^{N}_{k=1} \hat{\alpha } \right)  } \right)  +\sum\nolimits^{N}_{k=1} \left( \hat{\alpha }_{k} -1\right)  \left[ \psi \left( \hat{\alpha }_{k} \right)  -\psi \left( \sum\nolimits^{N}_{k-1} \hat{\alpha }_{k} \right)  \right],
    \label{LKL}
\end{equation}
where $\hat{\alpha } =y+\left( 1-y\right)  \odot \alpha$ is the adjusted parameter of the Dirichlet distribution which could avoid penalizing the evidence of the ground-truth class to 0, and $\Gamma \left( \right)$ is the gamma function.

In general, as shown in Eq.~\ref{lem} and Eq.~\ref{Ldl} to Eq.~\ref{LKL}, by introducing higher penalties for uncertain image-txt pairs via uncertainty calibration. For instance, for the text-image feature alignment with high uncertainty, it imposes an extra penalty to avoid the model focusing excessively on incorrect matches during contrastive pre-training.

We performed a comprehensive set of ablation studies (See Supplementary Tables 21–24) to elucidate the individual contributions of each module within RetiZero. The results confirm that our RetiZero model consistently surpasses all other configurations, illustrating the synergistic advantages gained from integrating an MAE-based encoder, text-image contrastive learning, and uncertainty calibration. Furthermore, we conducted additional ablation experiments using only publicly available datasets for pretraining and compared these outcomes to those of our full RetiZero model, which also incorporates data from ophthalmic literature and online resources (See Supplementary Table 25). Although public datasets alone are sufficient for reliably identifying common fundus diseases, they prove inadequate for robustly detecting rare conditions; the inclusion of more diverse data sources is thus indispensable. This finding underscores the critical importance of aggregating varied datasets to develop a knowledge-rich vision-language model with strong clinical applicability across a wide range of diseases.

\subsection{Definition of Dirichlet distribution}
The Dirichlet distribution was parameterized by its concentration K parameters $\bm{\alpha} =\left[ \alpha_{1} ,...,\alpha_{k} \right]$~\cite{ng2011dirichlet,han2020trusted}. Therefore, the probability density function of the Dirichlet distribution was computed as:
\begin{equation}
    D\left( P|\alpha \right)  =\begin{cases}\frac{1}{\beta \left( \alpha \right)  } \prod\nolimits^{K}_{k=1} p^{\alpha_{k} -1}_{k}&for\  P\in S_{K}\\ 0&Otherwise\end{cases},
\end{equation}
where $S_{K}$ was the $K$-dimensional unit simplex:
\begin{equation}
    S_{K}=\left\{ P|\sum\nolimits^{K}_{k=1} p_{k}=1\right\}  ,0\leq p_{k}\leq 1,
\end{equation}
where $\beta \left( \alpha \right)$ represented the $K$-dimensional multinomial beta function.

\subsection{AI-assisted clinical diagnosis settings:} To assess the capability of RetiZero's in recognizing fundus disease recognition without retraining the model, we randomly selected two images from each category of the EYE-52 dataset, creating a subset named EYE52-sub, which containing a total of 104 fundus photographs. We invited 19 ophthalmologists from 12 different institutions and hospitals across Singapore, the United States, and China, to make clinical diagnoses based on these 104 retinal images. Among the participating ophthalmologists, seven have 3 to 5 years of clinical experience, seven have 5 to 10 years, and five have more than 10 years. We developed an online fundus image reading system and uploaded the 104 images to the server (Fig. 3). To mimic the zero-shot setup, we provided 52 disease options as prompts on the webpage. During the image reading process, the clinicians selected a diagnosis from these 52 disease categories based solely on the image content. Additionally, each of the ophthalmologists was also asked to rate their confidence in their diagnostic results from level 1 to level 5. 

To evaluate whether our RetiZero can assist clinicians in improving their accuracy in diagnosing fundus disease, we conducted a second round of clinical evaluations with the same 19 ophthalmologists, one week after the initial round. We used the same set of retinal images and questions from the first round, but randomized the sequence of the images presented. For each question, RetiZero provided its top five prediction results as references. This approach allows for a comparative analysis of diagnostic performance with and without the assistance of RetiZero's predictions. Moreover, to mitigate potential bias, we conducted additional experiments using a new set of 104 images for the third round of testing to further demonstrate that our RetiZero can assist clinicians in improving their accuracy in diagnosing fundus disease (See Supplementary Table 26). 

We also conducted an in-depth analysis of the correlation between the position of the correct diagnosis within RetiZero's top five reference answers and the modifications made by ophthalmologists to their diagnoses. The top five results provided by RetiZero were scored as follows: correct diagnoses appearing in positions 1, 2, 3, 4, and 5 were assigned 5, 4, 3, 2, and 1 points, respectively, while those not appearing in the top five were assigned 0 points. This scoring system allowed us to calculate the top-ranking score for each case and assess how the ranking of the correct diagnosis with in the top five influenced diagnostic accuracy of the ophthalmologists. Moreover, we assessed the modifications made by ophthalmologists to their answers. Assuming an ophthalmologist's answers in the first and second rounds for the same case were denoted as (x, y), the scoring method was as follows: (True, True) = 0, (False, False) = 0, (True, False) = -1, (False, True) = 1. By analyzing the modification behaviour of 19 ophthalmologists, we calculated a response modification score for each case, which reflected the ophthalmologists’ thought process and decision-making during the diagnostic process.


\section{Code Availability}
The code is available at \url{https://github.com/LooKing9218/RetiZero}. 

\section{Data Availability}
The publicly available datasets used for pre-training are available at the following links and references:

\noindent APTOS: \url{https://www.kaggle.com/c/aptos2019-blindness-detection}. 

\noindent Cataract: \url{https://www.kaggle.com/datasets/jr2ngb/cataractdataset}. 

\noindent DDR: \url{https://github.com/nkicsl/DDR-dataset}. 

\noindent Diabetic Retinopathy Level Detection:\url{ https://www.kaggle.com/datasets/arbethi/diabetic-retinopathy-level-detection}.

\noindent Diabetic Retinopathy Organized: \url{https://www.kaggle.com/datasets/dola1507108/diabetic-retinopathy-organized}. 

\noindent DR15: \url{https://www.kaggle.com/datasets/nawa393/dr15_test}.

\noindent Messidor: \url{https://paperswithcode.com/dataset/messidor-1}.

\noindent MURED: \url{https://www.kaggle.com/datasets/abhirampolisetti/multi-label-retinal-disease-mured-dataset}.

\noindent Retina Dataset: \url{https://www.kaggle.com/datasets/jr2ngb/cataractdataset}. 

\noindent Kaggle DR: \url{https://www.kaggle.com/c/diabetic-retinopathy-detection/data}. 

\noindent ODIR5K: \url{https://www.kaggle.com/datasets/andrewmvd/ocular-disease-recognition-odir5k}. 

\noindent AUS dataset: \url{https://www.kaggle.com/competitions/innovation-challenge-2019/data}. 
\noindent SUSTech dataset~\cite{lin2020sustech}:
\url{ https://www.kaggle.com/datasets/mariaherrerot/the-sustechsysu-dataset}.

\noindent SiMES~\cite{foong2007rationale}, SIDNI~\cite{sabanayagam2017singapore}, ACRIMA~\cite{ACRIMA}, BEH~\cite{BEH}, DeepDRiD~\cite{DeepDRiD}, DR1-2~\cite{DR12}, E-ophta~\cite{Eophta}, AIROGS~\cite{AIROGS}, DeepEyeNet~\cite{DeepEyeNet}, FIVES~\cite{FIVES}, G1020~\cite{G1020}, Glaucoma dataset~\cite{Glaucoma_dataset,Glaucoma_dataset_1}, IDRiD~\cite{IDRiD}, JICHI~\cite{JICHI}, REFUGE~\cite{REFUGE}, ORIGA~\cite{ORIGA}, PARAGUAY~\cite{PARAGUAY}, EyePACS AirDoc~\cite{EyePACS_AirDoc}, JSIEC~\cite{cen2021automatic}, RFMid~\cite{RFMiD}.


\section{Supplementary file}
Supplementary file can be available at: \url{https://drive.google.com/file/d/1OmRagn47vLmK0WhX_hKTFT_qRYojRaos}

\bibliographystyle{IEEEtran}
\bibliography{references}

\begin{thebibliography}{10}
\providecommand{\url}[1]{#1}
\csname url@samestyle\endcsname
\providecommand{\newblock}{\relax}
\providecommand{\bibinfo}[2]{#2}
\providecommand{\BIBentrySTDinterwordspacing}{\spaceskip=0pt\relax}
\providecommand{\BIBentryALTinterwordstretchfactor}{4}
\providecommand{\BIBentryALTinterwordspacing}{\spaceskip=\fontdimen2\font plus
\BIBentryALTinterwordstretchfactor\fontdimen3\font minus \fontdimen4\font\relax}
\providecommand{\BIBforeignlanguage}[2]{{%
\expandafter\ifx\csname l@#1\endcsname\relax
\typeout{** WARNING: IEEEtran.bst: No hyphenation pattern has been}%
\typeout{** loaded for the language `#1'. Using the pattern for}%
\typeout{** the default language instead.}%
\else
\language=\csname l@#1\endcsname
\fi
#2}}
\providecommand{\BIBdecl}{\relax}
\BIBdecl

\bibitem{bellemo2019artificial}
V.~Bellemo, Z.~W. Lim, G.~Lim, Q.~D. Nguyen, Y.~Xie, M.~Y. Yip, H.~Hamzah, J.~Ho, X.~Q. Lee, W.~Hsu \emph{et~al.}, ``Artificial intelligence using deep learning to screen for referable and vision-threatening diabetic retinopathy in africa: a clinical validation study,'' \emph{The Lancet Digital Health}, vol.~1, no.~1, pp. e35--e44, 2019.

\bibitem{xie2020artificial}
Y.~Xie, Q.~D. Nguyen, H.~Hamzah, G.~Lim, V.~Bellemo, D.~V. Gunasekeran, M.~Y. Yip, X.~Q. Lee, W.~Hsu, M.~L. Lee \emph{et~al.}, ``Artificial intelligence for teleophthalmology-based diabetic retinopathy screening in a national programme: an economic analysis modelling study,'' \emph{The Lancet Digital Health}, vol.~2, no.~5, pp. e240--e249, 2020.

\bibitem{liu2019development}
H.~Liu, L.~Li, I.~M. Wormstone, C.~Qiao, C.~Zhang, P.~Liu, S.~Li, H.~Wang, D.~Mou, R.~Pang \emph{et~al.}, ``Development and validation of a deep learning system to detect glaucomatous optic neuropathy using fundus photographs,'' \emph{JAMA ophthalmology}, vol. 137, no.~12, pp. 1353--1360, 2019.

\bibitem{wang2020characterization}
M.~Wang, J.~Tichelaar, L.~R. Pasquale, L.~Q. Shen, M.~V. Boland, S.~R. Wellik, C.~G. De~Moraes, J.~S. Myers, P.~Ramulu, M.~Kwon \emph{et~al.}, ``Characterization of central visual field loss in end-stage glaucoma by unsupervised artificial intelligence,'' \emph{JAMA ophthalmology}, vol. 138, no.~2, pp. 190--198, 2020.

\bibitem{peng2021automatic}
Y.~Peng, W.~Zhu, Z.~Chen, M.~Wang, L.~Geng, K.~Yu, Y.~Zhou, T.~Wang, D.~Xiang, F.~Chen \emph{et~al.}, ``Automatic staging for retinopathy of prematurity with deep feature fusion and ordinal classification strategy,'' \emph{IEEE Transactions on Medical Imaging}, vol.~40, no.~7, pp. 1750--1762, 2021.

\bibitem{taylor2019monitoring}
S.~Taylor, J.~M. Brown, K.~Gupta, J.~P. Campbell, S.~Ostmo, R.~P. Chan, J.~Dy, D.~Erdogmus, S.~Ioannidis, S.~J. Kim \emph{et~al.}, ``Monitoring disease progression with a quantitative severity scale for retinopathy of prematurity using deep learning,'' \emph{JAMA ophthalmology}, vol. 137, no.~9, pp. 1022--1028, 2019.

\bibitem{de2018clinically}
J.~De~Fauw, J.~R. Ledsam, B.~Romera-Paredes, S.~Nikolov, N.~Tomasev, S.~Blackwell, H.~Askham, X.~Glorot, B.~O’Donoghue, D.~Visentin \emph{et~al.}, ``Clinically applicable deep learning for diagnosis and referral in retinal disease,'' \emph{Nature medicine}, vol.~24, no.~9, pp. 1342--1350, 2018.

\bibitem{wang2023uncertainty}
M.~Wang, T.~Lin, L.~Wang, A.~Lin, K.~Zou, X.~Xu, Y.~Zhou, Y.~Peng, Q.~Meng, Y.~Qian \emph{et~al.}, ``Uncertainty-inspired open set learning for retinal anomaly identification,'' \emph{Nature Communications}, vol.~14, no.~1, p. 6757, 2023.

\bibitem{cen2021automatic}
L.-P. Cen, J.~Ji, J.-W. Lin, S.-T. Ju, H.-J. Lin, T.-P. Li, Y.~Wang, J.-F. Yang, Y.-F. Liu, S.~Tan \emph{et~al.}, ``Automatic detection of 39 fundus diseases and conditions in retinal photographs using deep neural networks,'' \emph{Nature communications}, vol.~12, no.~1, p. 4828, 2021.

\bibitem{huang2023visual}
Z.~Huang, F.~Bianchi, M.~Yuksekgonul, T.~J. Montine, and J.~Zou, ``A visual--language foundation model for pathology image analysis using medical twitter,'' \emph{Nature medicine}, vol.~29, no.~9, pp. 2307--2316, 2023.

\bibitem{radford2021learning}
A.~Radford, J.~W. Kim, C.~Hallacy, A.~Ramesh, G.~Goh, S.~Agarwal, G.~Sastry, A.~Askell, P.~Mishkin, J.~Clark \emph{et~al.}, ``Learning transferable visual models from natural language supervision,'' in \emph{International conference on machine learning}.\hskip 1em plus 0.5em minus 0.4em\relax PMLR, 2021, pp. 8748--8763.

\bibitem{zhang2020asymmetric}
M.~Zhang, S.~X. Fei, J.~Liu, S.~Xu, Y.~Piao, and H.~Lu, ``Asymmetric two-stream architecture for accurate rgb-d saliency detection,'' in \emph{Computer Vision--ECCV 2020: 16th European Conference, Glasgow, UK, August 23--28, 2020, Proceedings, Part XXVIII 16}.\hskip 1em plus 0.5em minus 0.4em\relax Springer, 2020, pp. 374--390.

\bibitem{zhang2022few}
Z.~Zhang, Y.~Li, Q.~Zhai, Y.~Li, and M.~Gao, ``Few-shot learning for fine-grained signal modulation recognition based on foreground segmentation,'' \emph{IEEE Transactions on Vehicular Technology}, vol.~71, no.~3, pp. 2281--2292, 2022.

\bibitem{lai2024carzero}
H.~Lai, Q.~Yao, Z.~Jiang, R.~Wang, Z.~He, X.~Tao, and S.~K. Zhou, ``Carzero: Cross-attention alignment for radiology zero-shot classification,'' in \emph{IEEE Conf. Comput. Vis. Pattern Recog.}, 2024.

\bibitem{RETFound}
Y.~Zhou, M.~A. Chia, S.~K. Wagner, M.~S. Ayhan, D.~J. Williamson, R.~R. Struyven, T.~Liu, M.~Xu, M.~G. Lozano, P.~Woodward-Court \emph{et~al.}, ``A foundation model for generalizable disease detection from retinal images,'' \emph{Nature}, vol. 622, no. 7981, pp. 156--163, 2023.

\bibitem{he2022masked}
K.~He, X.~Chen, S.~Xie, Y.~Li, P.~Doll{\'a}r, and R.~Girshick, ``Masked autoencoders are scalable vision learners,'' in \emph{Proceedings of the IEEE/CVF conference on computer vision and pattern recognition}, 2022, pp. 16\,000--16\,009.

\bibitem{FLAIR}
J.~Silva-Rodriguez, H.~Chakor, R.~Kobbi, J.~Dolz, and I.~B. Ayed, ``A foundation language-image model of the retina (flair): Encoding expert knowledge in text supervision,'' \emph{arXiv preprint arXiv:2308.07898}, 2023.

\bibitem{selvaraju2017grad}
R.~R. Selvaraju, M.~Cogswell, A.~Das, R.~Vedantam, D.~Parikh, and D.~Batra, ``Grad-cam: Visual explanations from deep networks via gradient-based localization,'' in \emph{Proceedings of the IEEE international conference on computer vision}, 2017, pp. 618--626.

\bibitem{quellec2020automatic}
G.~Quellec, M.~Lamard, P.-H. Conze, P.~Massin, and B.~Cochener, ``Automatic detection of rare pathologies in fundus photographs using few-shot learning,'' \emph{Medical image analysis}, vol.~61, p. 101660, 2020.

\bibitem{gao2023discriminative}
M.~Gao, H.~Jiang, L.~Zhu, Z.~Jiang, M.~Geng, Q.~Ren, and Y.~Lu, ``Discriminative ensemble meta-learning with co-regularization for rare fundus diseases diagnosis,'' \emph{Medical Image Analysis}, vol.~89, p. 102884, 2023.

\bibitem{yi2023label}
D.~Yi, Y.~Hua, P.~Murchie, and P.~K. Sharma, ``Label-free medical image quality evaluation by semantics-aware contrastive learning in iomt,'' \emph{IEEE journal of biomedical and health informatics}, 2023.

\bibitem{mahapatra2022self}
D.~Mahapatra, Z.~Ge, and M.~Reyes, ``Self-supervised generalized zero shot learning for medical image classification using novel interpretable saliency maps,'' \emph{IEEE Transactions on Medical Imaging}, vol.~41, no.~9, pp. 2443--2456, 2022.

\bibitem{fang2021deep}
J.~Fang, H.~Fu, and J.~Liu, ``Deep triplet hashing network for case-based medical image retrieval,'' \emph{Medical image analysis}, vol.~69, p. 101981, 2021.

\bibitem{fang2023deep}
J.~Fang, M.~Zeng, X.~Zhang, H.~Liu, Y.~Zhao, P.~Zhang, H.~Yang, J.~Liu, H.~Miao, Y.~Hu \emph{et~al.}, ``Deep metric learning with mirror attention and fine triplet loss for fundus image retrieval in ophthalmology,'' \emph{Biomedical Signal Processing and Control}, vol.~80, p. 104277, 2023.

\bibitem{chandakkar2012machine}
P.~S. Chandakkar, R.~Venkatesan, B.~Li, and H.~K. Li, ``A machine-learning approach to retrieving diabetic retinopathy images,'' in \emph{Proceedings of the ACM Conference on Bioinformatics, Computational Biology and Biomedicine}, 2012, pp. 588--589.

\bibitem{quellec2011automated}
G.~Quellec, M.~Lamard, G.~Cazuguel, L.~Bekri, W.~Daccache, C.~Roux, and B.~Cochener, ``Automated assessment of diabetic retinopathy severity using content-based image retrieval in multimodal fundus photographs,'' \emph{Investigative ophthalmology \& visual science}, vol.~52, no.~11, pp. 8342--8348, 2011.

\bibitem{dong2022artificial}
L.~Dong, W.~He, R.~Zhang, Z.~Ge, Y.~X. Wang, J.~Zhou, J.~Xu, L.~Shao, Q.~Wang, Y.~Yan \emph{et~al.}, ``Artificial intelligence for screening of multiple retinal and optic nerve diseases,'' \emph{JAMA network open}, vol.~5, no.~5, pp. e229\,960--e229\,960, 2022.

\bibitem{ruamviboonsuk2022real}
P.~Ruamviboonsuk, R.~Tiwari, R.~Sayres, V.~Nganthavee, K.~Hemarat, A.~Kongprayoon, R.~Raman, B.~Levinstein, Y.~Liu, M.~Schaekermann \emph{et~al.}, ``Real-time diabetic retinopathy screening by deep learning in a multisite national screening programme: a prospective interventional cohort study,'' \emph{The Lancet Digital Health}, vol.~4, no.~4, pp. e235--e244, 2022.

\bibitem{li2024performance}
B.~Li, H.~Chen, W.~Yu, M.~Zhang, F.~Lu, J.~Ma, Y.~Hao, X.~Li, B.~Hu, L.~Shen \emph{et~al.}, ``The performance of a deep learning system in assisting junior ophthalmologists in diagnosing 13 major fundus diseases: a prospective multi-center clinical trial,'' \emph{NPJ digital medicine}, vol.~7, no.~1, p.~8, 2024.

\bibitem{lin2021application}
D.~Lin, J.~Xiong, C.~Liu, L.~Zhao, Z.~Li, S.~Yu, X.~Wu, Z.~Ge, X.~Hu, B.~Wang \emph{et~al.}, ``Application of comprehensive artificial intelligence retinal expert (care) system: a national real-world evidence study,'' \emph{The Lancet Digital Health}, vol.~3, no.~8, pp. e486--e495, 2021.

\bibitem{li2022development}
B.~Li, H.~Chen, B.~Zhang, M.~Yuan, X.~Jin, B.~Lei, J.~Xu, W.~Gu, D.~C.~S. Wong, X.~He \emph{et~al.}, ``Development and evaluation of a deep learning model for the detection of multiple fundus diseases based on colour fundus photography,'' \emph{British Journal of Ophthalmology}, vol. 106, no.~8, pp. 1079--1086, 2022.

\bibitem{gui2021review}
J.~Gui, Z.~Sun, Y.~Wen, D.~Tao, and J.~Ye, ``A review on generative adversarial networks: Algorithms, theory, and applications,'' \emph{IEEE transactions on knowledge and data engineering}, vol.~35, no.~4, pp. 3313--3332, 2021.

\bibitem{rombach2022high}
R.~Rombach, A.~Blattmann, D.~Lorenz, P.~Esser, and B.~Ommer, ``High-resolution image synthesis with latent diffusion models,'' in \emph{Proceedings of the IEEE/CVF conference on computer vision and pattern recognition}, 2022, pp. 10\,684--10\,695.

\bibitem{shorten2019survey}
C.~Shorten and T.~M. Khoshgoftaar, ``A survey on image data augmentation for deep learning,'' \emph{Journal of big data}, vol.~6, no.~1, pp. 1--48, 2019.

\bibitem{hu2021lora}
E.~J. Hu, Y.~Shen, P.~Wallis, Z.~Allen-Zhu, Y.~Li, S.~Wang, L.~Wang, and W.~Chen, ``Lora: Low-rank adaptation of large language models,'' \emph{arXiv preprint arXiv:2106.09685}, 2021.

\bibitem{jie2023fact}
S.~Jie and Z.-H. Deng, ``Fact: Factor-tuning for lightweight adaptation on vision transformer,'' in \emph{Proceedings of the AAAI conference on artificial intelligence}, vol.~37, no.~1, 2023, pp. 1060--1068.

\bibitem{chen2024ma}
C.~Chen, J.~Miao, D.~Wu, A.~Zhong, Z.~Yan, S.~Kim, J.~Hu, Z.~Liu, L.~Sun, X.~Li \emph{et~al.}, ``Ma-sam: Modality-agnostic sam adaptation for 3d medical image segmentation,'' \emph{Medical Image Analysis}, vol.~98, p. 103310, 2024.

\bibitem{alsentzer2019publicly}
E.~Alsentzer, J.~R. Murphy, W.~Boag, W.-H. Weng, D.~Jin, T.~Naumann, and M.~McDermott, ``Publicly available clinical bert embeddings,'' \emph{arXiv preprint arXiv:1904.03323}, 2019.

\bibitem{han2020trusted}
Z.~Han, C.~Zhang, H.~Fu, and J.~T. Zhou, ``Trusted multi-view classification,'' in \emph{International Conference on Learning Representations}, 2020.

\bibitem{ng2011dirichlet}
K.~W. Ng, G.-L. Tian, and M.-L. Tang, ``Dirichlet and related distributions: Theory, methods and applications,'' 2011.

\bibitem{lin2020sustech}
L.~Lin, M.~Li, Y.~Huang, P.~Cheng, H.~Xia, K.~Wang, J.~Yuan, and X.~Tang, ``The sustech-sysu dataset for automated exudate detection and diabetic retinopathy grading,'' \emph{Scientific Data}, vol.~7, no.~1, p. 409, 2020.

\bibitem{foong2007rationale}
A.~W. Foong, S.-M. Saw, J.-L. Loo, S.~Shen, S.-C. Loon, M.~Rosman, T.~Aung, D.~T. Tan, E.~S. Tai, and T.~Y. Wong, ``Rationale and methodology for a population-based study of eye diseases in malay people: The singapore malay eye study (simes),'' \emph{Ophthalmic epidemiology}, vol.~14, no.~1, pp. 25--35, 2007.

\bibitem{sabanayagam2017singapore}
C.~Sabanayagam, W.~Yip, P.~Gupta, R.~B. Mohd~Abdul, E.~Lamoureux, N.~Kumari, G.~C. Cheung, C.~Y. Cheung, J.~J. Wang, C.-Y. Cheng \emph{et~al.}, ``Singapore indian eye study-2: methodology and impact of migration on systemic and eye outcomes,'' \emph{Clinical \& experimental ophthalmology}, vol.~45, no.~8, pp. 779--789, 2017.

\bibitem{ACRIMA}
A.~Diaz-Pinto, S.~Morales, V.~Naranjo, T.~K{\"o}hler, J.~M. Mossi, and A.~Navea, ``Cnns for automatic glaucoma assessment using fundus images: an extensive validation,'' \emph{Biomedical engineering online}, vol.~18, pp. 1--19, 2019.

\bibitem{BEH}
M.~T. Islam, S.~T. Mashfu, A.~Faisal, S.~C. Siam, I.~T. Naheen, and R.~Khan, ``Deep learning-based glaucoma detection with cropped optic cup and disc and blood vessel segmentation,'' \emph{Ieee Access}, vol.~10, pp. 2828--2841, 2021.

\bibitem{DeepDRiD}
R.~Liu, X.~Wang, Q.~Wu, L.~Dai, X.~Fang, T.~Yan, J.~Son, S.~Tang, J.~Li, Z.~Gao \emph{et~al.}, ``Deepdrid: Diabetic retinopathy—grading and image quality estimation challenge,'' \emph{Patterns}, vol.~3, no.~6, 2022.

\bibitem{DR12}
R.~Pires, H.~F. Jelinek, J.~Wainer, E.~Valle, and A.~Rocha, ``Advancing bag-of-visual-words representations for lesion classification in retinal images,'' \emph{PloS one}, vol.~9, no.~6, p. e96814, 2014.

\bibitem{Eophta}
E.~Decenciere, G.~Cazuguel, X.~Zhang, G.~Thibault, J.-C. Klein, F.~Meyer, B.~Marcotegui, G.~Quellec, M.~Lamard, R.~Danno \emph{et~al.}, ``Teleophta: Machine learning and image processing methods for teleophthalmology,'' \emph{Irbm}, vol.~34, no.~2, pp. 196--203, 2013.

\bibitem{AIROGS}
C.~De~Vente, K.~A. Vermeer, N.~Jaccard, H.~Wang, H.~Sun, F.~Khader, D.~Truhn, T.~Aimyshev, Y.~Zhanibekuly, T.-D. Le \emph{et~al.}, ``Airogs: artificial intelligence for robust glaucoma screening challenge,'' \emph{IEEE transactions on medical imaging}, 2023.

\bibitem{DeepEyeNet}
J.-H. Huang, C.-H.~H. Yang, F.~Liu, M.~Tian, Y.-C. Liu, T.-W. Wu, I.~Lin, K.~Wang, H.~Morikawa, H.~Chang \emph{et~al.}, ``Deepopht: medical report generation for retinal images via deep models and visual explanation,'' in \emph{Proceedings of the IEEE/CVF winter conference on applications of computer vision}, 2021, pp. 2442--2452.

\bibitem{FIVES}
K.~Jin, X.~Huang, J.~Zhou, Y.~Li, Y.~Yan, Y.~Sun, Q.~Zhang, Y.~Wang, and J.~Ye, ``Fives: A fundus image dataset for artificial intelligence based vessel segmentation,'' \emph{Scientific Data}, vol.~9, no.~1, p. 475, 2022.

\bibitem{G1020}
M.~N. Bajwa, G.~A.~P. Singh, W.~Neumeier, M.~I. Malik, A.~Dengel, and S.~Ahmed, ``G1020: A benchmark retinal fundus image dataset for computer-aided glaucoma detection,'' in \emph{2020 International Joint Conference on Neural Networks (IJCNN)}.\hskip 1em plus 0.5em minus 0.4em\relax IEEE, 2020, pp. 1--7.

\bibitem{Glaucoma_dataset}
A.~Singh, M.~K. Dutta, M.~ParthaSarathi, V.~Uher, and R.~Burget, ``Image processing based automatic diagnosis of glaucoma using wavelet features of segmented optic disc from fundus image,'' \emph{Computer methods and programs in biomedicine}, vol. 124, pp. 108--120, 2016.

\bibitem{Glaucoma_dataset_1}
A.~Issac, M.~P. Sarathi, and M.~K. Dutta, ``An adaptive threshold based image processing technique for improved glaucoma detection and classification,'' \emph{Computer methods and programs in biomedicine}, vol. 122, no.~2, pp. 229--244, 2015.

\bibitem{IDRiD}
P.~Porwal, S.~Pachade, M.~Kokare, G.~Deshmukh, J.~Son, W.~Bae, L.~Liu, J.~Wang, X.~Liu, L.~Gao \emph{et~al.}, ``Idrid: Diabetic retinopathy--segmentation and grading challenge,'' \emph{Medical image analysis}, vol.~59, p. 101561, 2020.

\bibitem{JICHI}
H.~Takahashi, H.~Tampo, Y.~Arai, Y.~Inoue, and H.~Kawashima, ``Applying artificial intelligence to disease staging: Deep learning for improved staging of diabetic retinopathy,'' \emph{PloS one}, vol.~12, no.~6, p. e0179790, 2017.

\bibitem{REFUGE}
J.~I. Orlando, H.~Fu, J.~B. Breda, K.~Van~Keer, D.~R. Bathula, A.~Diaz-Pinto, R.~Fang, P.-A. Heng, J.~Kim, J.~Lee \emph{et~al.}, ``Refuge challenge: A unified framework for evaluating automated methods for glaucoma assessment from fundus photographs,'' \emph{Medical image analysis}, vol.~59, p. 101570, 2020.

\bibitem{ORIGA}
Z.~Zhang, F.~S. Yin, J.~Liu, W.~K. Wong, N.~M. Tan, B.~H. Lee, J.~Cheng, and T.~Y. Wong, ``Origa-light: An online retinal fundus image database for glaucoma analysis and research,'' in \emph{2010 Annual international conference of the IEEE engineering in medicine and biology}.\hskip 1em plus 0.5em minus 0.4em\relax IEEE, 2010, pp. 3065--3068.

\bibitem{PARAGUAY}
V.~E.~C. Ben{\'\i}tez, I.~C. Matto, J.~C.~M. Rom{\'a}n, J.~L.~V. Noguera, M.~Garc{\'\i}a-Torres, J.~Ayala, D.~P. Pinto-Roa, P.~E. Gardel-Sotomayor, J.~Facon, and S.~A. Grillo, ``Dataset from fundus images for the study of diabetic retinopathy,'' \emph{Data in brief}, vol.~36, p. 107068, 2021.

\bibitem{EyePACS_AirDoc}
L.~Ju, X.~Wang, L.~Wang, D.~Mahapatra, X.~Zhao, Q.~Zhou, T.~Liu, and Z.~Ge, ``Improving medical images classification with label noise using dual-uncertainty estimation,'' \emph{IEEE transactions on medical imaging}, vol.~41, no.~6, pp. 1533--1546, 2022.

\bibitem{RFMiD}
S.~Pachade, P.~Porwal, D.~Thulkar, M.~Kokare, G.~Deshmukh, V.~Sahasrabuddhe, L.~Giancardo, G.~Quellec, and F.~M{\'e}riaudeau, ``Retinal fundus multi-disease image dataset (rfmid): A dataset for multi-disease detection research,'' \emph{Data}, vol.~6, no.~2, p.~14, 2021.

\end{thebibliography}

\clearpage
\setcounter{table}{0} 
\setcounter{figure}{0} 
\captionsetup[figure]{labelfont={bf},name={Supplementary Figure: },labelsep=period}
\captionsetup[table]{labelfont={bf},name={Supplementary Table: },labelsep=period}
\newcolumntype{L}{>{\hspace{0.1in}\arraybackslash}p{0.4\textwidth}}
\newcolumntype{C}{>{\centering\arraybackslash}p{0.4\textwidth}}

\end{document}